\documentclass[12pt]{article}
\usepackage{amsmath}
\usepackage{amssymb}
\usepackage{myart}

\input tcilatex
\begin{document}

\title{Dirac Equation in Terms of Hydrodynamic Variables}
\author{Yuri A.Rylov}
\date{Institute for Problems in Mechanics, Russian Academy of Science,\\
101 Vernadskii Ave., Moscow, 117526, Russia.}

\centerline{\textbf{ Comment for physicists, written 30 January
2011}}

\bigskip

This paper has been published in 1995 in the mathematical journal \textit{%
Advances in Applied Clifford Algebras.} It was my unique
publication on Clifford algebras. Nevertheless I was personally
invited to participate in several conferences on Clifford
algebras, although I am a physicist, and a use of Clifford
algebras is only a mathematical method of the Dirac equation
representation. In my paper the $\gamma $-matrices are eliminated,
and the
Dirac equation is presented in the hydrodynamical form, where the $\gamma $%
-matrices are absent. On one hand, the physicists do not use the
Sauter - Sommerfeld method of the $\gamma $-matrices
representation in the form of Clifford algebras. On the other
hand, one can eliminate $\gamma $-matrices, only if they are not
represented in a concrete representation (i.e. they are to be
considered as basic elements of a Clifford algebra).

Elimination of $\gamma $-matrices admits one to obtain essential
physical results: (1) discovery that the Dirac particle has a
complicate internal structure (it is not a pointlike particle),
(2) discovery that the internal degrees of freedom are described
non-relativistically. These statements are very essential, but
they are rather unexpected simultaneously, and one cannot
understand them without a knowledge of the Sauter - Sommerfeld
method. I tried to publish the physically essential parts of the paper \cite%
{R2004,R2004a} in physical reviewed journals. Unfortunately, I
failed, because, apparently, the reviewers did not know the Sauter
- Sommerfeld method of the $\gamma $-matrices representation. They
do not trust in mathematical correctness of my results, although
experts in Clifford algebras, publishing me paper in the
mathematical journal did.

\maketitle

\begin{abstract}
The distributed system $\mathcal{S}_D$ described by the Dirac equation is
investigated simply as a dynamic system, i.e. without usage of quantum
principles. The Dirac equation is described in terms of hydrodynamic
variables: 4-flux $j^{i}$, pseudo-vector of the spin $S^{i}$, an action $%
\hbar \varphi $ and a pseudo-scalar $\kappa $. In the quasi-uniform
approximation, when all transversal derivatives (orthogonal to the flux
vector $j^i$) are small, the system $\mathcal{S}_D$ turns to a statistical
ensemble of classical concentrated systems $\mathcal{S}_{dc}$. Under some
conditions the classical system $\mathcal{S}_{dc}$ describes a classical
pointlike particle moving in a given electromagnetic field. In general, the
world line of the particle is a helix, even if the electromagnetic field is
absent. Both dynamic systems $\mathcal{S}_D$ and $\mathcal{S}_{dc}$ appear
to be non-relativistic in the sense that the dynamic equations written in
terms of hydrodynamic variables are not relativistically covariant with
respect to them, although all dynamic variables are tensors or
pseudo-tensors. They becomes relativistically covariant only after addition
of a constant unit timelike vector $f^{i}$ which should be considered as a
dynamic variable describing a space-time property. This "constant" variable
arises instead of $\gamma $-matrices which are removed by means of zero
divizors in the course of the transformation to hydrodynamic variables. It
is possible to separate out dynamic variables $\kappa $, $\kappa ^i$
responsible for quantum effects. It means that, setting $\kappa ,\kappa
^i\equiv 0$, the dynamic system $\mathcal{S}_D$ described by the Dirac
equation turns to a statistical ensemble $\mathcal{E}_{Dqu}$ of classical
dynamic systems $\mathcal{S}_{dc}$.
\end{abstract}

\section{Introduction}

The Dirac equation considered as a dynamic equation for a wave function in
frames of quantum mechanics principles has been investigated almost
completely, and it is hardly possible to add anything new. At the same time
the Dirac equation considered as a dynamic equation for a distributed
dynamic system $\mathcal{S}_{D}$ displays a series of such unexpected
properties as existence of dynamic variables $\kappa $ responsible for
quantum effects and appearance of a constant timelike vector describing a
split of the space-time into the space and the time.

From mathematical point of view our investigation of the Dirac equation is
simply a change of variables, when the Dirac four-component complex wave
function is substituted by certain tensor variables having also eight
independent real components. The tensor variables are: the 4-flux vector $%
j^{i}$, $i=0,1,2,3$, the spin pseudo-vector $S^{i}$, $i=0,1,2,3$, the action
scalar $\varphi $, and a pseudo-scalar $\kappa $. These quantities will be
referred to as hydrodynamic variables.

From physical standpoint the dynamic system $\mathcal{S}_{D}$ is considered
as a distributed dynamic system describing in some way an electron (or
positron) motion. The motion of a single particle is supposed to be
stochastic, the variables $j^{i}$ describing the mean 4-flux of particles.
Interpretation of other dynamic variables of $\mathcal{S}_{D}$ is produced
on the base of a comparison with a statistical ensemble of classical
particles. The quantal correspondence principle, when a linear operator
corresponds to any physical quantity, is not used. Other quantum principles
are not used also. Instead of them one uses a more general statistical
principle which asserts: \textit{A set} $\mathcal{E}$ (statistical ensemble)
\textit{of many similar independent stochastic systems} $\mathcal{S}_{s}$
\textit{is a deterministic dynamic system} $\mathcal{S}_{d}$ [1]. The term
"a stochastic system" means that experiments with a single system are
irreproducible, and there are no dynamic equations for such a system. If $%
\mathcal{S}$ means either stochastic system $\mathcal{S}_{s}$, or a
deterministic system $\mathcal{S}_{d}$, the statistical principle can be
formulated in the form
\begin{equation*}
\mathcal{E}[\mathcal{S}]\quad \text{is}\quad \mathcal{S}_{d}
\end{equation*}
The statistical principle can be applied both to quantum and classical
systems. It is a more general statement, than the set of quantum principles.
In particular, it can be valid for such stochastic systems which are neither
quantum, nor classical. The main concept of this approach is the statistical
ensemble considered as a dynamic system (not a probability density, or a
probability amplitude). This approach will be referred to as \textit{a
statistical ensemble technique} (SET).

There is a reason for consideration of the Dirac equation without usage of
quantum principles, i.e. simply as a dynamic equation for the system $%
\mathcal{S}_D$. Recently one suggested a hypothesis that the real space-time
is a distorted Minkowski space-time, and the distortion of the space-time is
a reason for quantum effects [2]. Distortion is such a deformation of the
Minkowski space-time which transform one-dimensional world lines into
three-dimensional world tubes. The world tubes of particles appear
stochastic, and their statistical description coincides with quantum
description provided the Planck constant $\hbar $ determines the space-time
distortion (thickness of the tubes). This hypothesis was proved for a
non-relativistic free particle. It is interesting to test the hypothesis in
the case of a relativistic particle. For such a test it is necessary to
separate the action describing the ensemble of Dirac particles into two
parts: a classical part and a part responsible for quantum effects.

Another motive is as follows. The conventional quantum dynamics technique
(QDT) is an axiomatic construction. QDT relates to SET approximately in the
same way, as the axiomatic thermodynamics relates to the statistical
physics. For instance, the Brownian motion cannot be explained and
understood from standpoint of thermodynamics. Thermodynamic axioms do not
permit to do this. The statistical physics explains the Brownian motion
phenomenon and restricts an application of thermodynamic principles, because
a statistical approach is more general, than the approach of the axiomatic
thermodynamics.

Something like that one can see in the field of the quantum dynamics.
Problem of pair production is a principal problem of the high energy
physics. In those areas, where the pair production is unessential and can be
considered as a correction (for instance, in quantum electrodynamics) the
quantum theory succeeds. In those areas, where pair production is a
dominating effect, the quantum theory (in particular, QFT) failed.

For last fifty years the quantum field theory has not succeeded in solving
the problem of pair production. A suspicion arises that this failure is not
accidental. Maybe, the problem of pair production cannot be solved in
framework of the quantum principles, and a more general statistical approach
is necessary.

Classical limit of the Dirac equation is rather difficult to obtain, because
it contains such non-classical quantities as Dirac $\gamma $-matrices, which
hardly can be considered from the classical point of view.

Usually a quantum electron moving in a given electromagnetic field $A_l$, $%
l=0,1,2,3$ is described by the Dirac equation
\begin{equation}
-i\hbar \gamma ^{l}\partial _{l}\psi +eA_{l}\gamma ^{l}\psi +m\psi =0,
\label{a1.1}
\end{equation}%
where $\psi $ is a four-component complex wave function. The light speed $c$
is chosen to be equal to 1. It is possible to transform the variables $\psi $
and to describe this system in terms of variables $\varphi ,j^{l},S^{l}$, $%
(l=0,1,2,3)$, $\kappa $, defined by the relations

\begin{equation*}
j^{l}=\bar{\psi}\gamma ^{l}\psi ,\qquad l=0,1,2,3,\qquad \bar{\psi}=\psi
^{\ast }\gamma ^{0};
\end{equation*}%
\begin{equation*}
S^{l}=i\bar{\psi}\gamma _{5}\gamma ^{l}\psi ,\qquad l=0,1,2,3,\qquad \gamma
_{5}=\gamma ^{0123}\equiv \gamma ^{0}\gamma ^{1}\gamma ^{2}\gamma ^{3};
\end{equation*}%
\begin{equation}
\partial _{l}\varphi =(\bar{\psi}\partial _{l}\psi -\partial _{l}\bar{\psi}%
\psi )(2i\bar{\psi}\psi )^{-1},\qquad l=0,1,2,3;  \label{a1.2}
\end{equation}%
\begin{equation*}
\cos \kappa =\bar{\psi}\psi (j^{l}j_{l})^{-1/2};
\end{equation*}%
Here $\gamma ^{l}$, $l=0,1,2,3$ are the Dirac $\gamma $-matrices satisfying
the commutation relation

\begin{equation}
\gamma ^{i}\gamma ^{k}+\gamma ^{k}\gamma ^{i}=2g^{ik},\qquad i,k=0,1,2,3,
\label{a1.3}
\end{equation}%
where $g^{ik}$ =diag$(1,-1,-1,-1)$ is the metric tensor. Only two components
of the pseudo-vector $S^{l}$ are independent, because there are two
identities

\begin{equation}
S^{l}S_{l}\equiv -j^{l}j_{l},\qquad j^{l}S_{l}\equiv 0.  \label{a1.4}
\end{equation}%
Pseudo-vector $S^{l}$ is treated as a spin pseudo-vector, because it is
connected uniquely with the spin tensor $S^{ml,k}$, defined by the relation
[3]

\begin{equation}
S^{ml,k}={\frac{1}{4}}\bar{\psi}(\gamma ^{k}\sigma ^{lm}+\sigma ^{lm}\gamma
^{k})\psi ,\qquad \sigma ^{lm}={\frac{i}{2}}(\gamma ^{l}\gamma ^{m}-\gamma
^{m}\gamma ^{l})  \label{a1.5}
\end{equation}%
The relation between $S^{i}$ and $S^{ml,k}$ has the form

\begin{equation*}
S^{i}={\frac{1}{3}}g^{ij}\varepsilon _{jmlk}S^{ml,k},\qquad i=0,1,2,3;
\end{equation*}%
\begin{equation}
S^{ml,k}=-{\frac{1}{2}}\varepsilon ^{imlk}S_{i},\qquad m,l,k=0,1,2,3.
\label{a1.6}
\end{equation}%
Here $\varepsilon _{jmlk}$ and $\varepsilon ^{imlk}$ are Levi-Chivita
pseudo-tensors $(\varepsilon _{0123}=1$, $\varepsilon ^{0123}=-1)$.

Thus, the scalar $\varphi $, pseudo-scalar $\kappa $, vector $j^{l}$, and
pseudo-vector $S^{l}$ have 8 real independent components which are used
instead of 8 real components (4 complex components) of the Dirac wave
function $\psi $.

Realization of such a transformation is a goal of the present paper. Such a
description can be regarded as a description in terms of hydrodynamic
variables. Under some conditions the dynamic system $\mathcal{S}_{D}$ turns
into a statistical ensemble $\mathcal{E}_{Dqu}$ of classical dynamic systems
$\mathcal{S}_{dc}$. The $\mathcal{S}_{dc}$ can be interpreted as a classical
analog of the Dirac electron. $\mathcal{E}_{Dqu}$ is described usually in
terms of hydrodynamic variables $j^{l}$, $\varphi $, $\mathbf{\xi }$. The
dynamic systems $\mathcal{S}_{D}$ and $\mathcal{E}_{Dqu}$ are distinguished
by some terms of their Lagrangians, but not conceptually. Thus, one can
consider a dynamic distinction between the Dirac dynamic system $\mathcal{S}%
_{D}$ and the corresponding classical ensemble.

In the second section some properties of the statistical ensemble $\mathcal{E%
}_{cl}$ are investigated. The third section is devoted to introduction of
hydrodynamic variables. In the fourth section the action for the Dirac
equation is written in terms of hydrodynamic variables. The fifth section is
devoted to consideration of quasi-uniform state of the system $\mathcal{S}_D$%
. Relativistic invariance of dynamic equation in terms of hydrodynamic
variables is discussed in the sixth section. The seventh section is devoted
to investigation of the classical analog $\mathcal{S}_{dc}$ of the Dirac
electron.

\section{Statistical ensemble of classical dynamic \newline
systems}

Let there be a classical system $\mathcal{S}$ described by the Lagrangian
function $L(t,\mathbf{x},d\mathbf{x}/dt)$, where $\mathbf{x}=\{x^{\alpha }\}$%
, $d\mathbf{x}/dt=\{dx^{\alpha }/dt\}$, $\alpha =1,2,\ldots n$ are
generalized coordinates and velocities. Then, by definition a pure
statistical ensemble $\mathcal{E}_{cl}$ of systems $\mathcal{S}$ is a set of
similar independent systems $\mathcal{S}$. Its Lagrangian is a sum
(integral) of Lagrangians $L$. The action for the ensemble $\mathcal{E}_{cl}$
has the form

\begin{equation}
\mathcal{A}_{L}[\mathbf{x}]=\int L(t,\mathbf{x},d\mathbf{x}/dt)dtd\mathbf{%
\xi },\qquad d\mathbf{\xi }=\prod\limits_{\alpha =1}^{\alpha =n}d\xi
_{\alpha }  \label{a2.1}
\end{equation}%
where $\mathbf{\xi }=\{\xi _{\alpha }\}$, $\alpha =1,2,\ldots n$ are
Lagrangian coordinates labelling the systems of the ensemble, and $\mathbf{x}%
=\mathbf{x}(t,\mathbf{\xi })$ is a function of $t$ and $\mathbf{\xi }$.

The dynamic system $\mathcal{E}_{cl}$ can be considered as a fluid described
in the Lagrangian coordinates (the time $t$ and Lagrangian coordinates $%
\mathbf{\xi }$ are independent variables). The same action written in the
Euler coordinates ($t,\mathbf{x}$ are independent variables) has the form
\begin{equation}
\mathcal{A}_{E}[j,\varphi ,\mathbf{\xi }]=\int \{\mathcal{L}(x,j)-\hbar
j^{i}[\partial _{i}\varphi +g^{\alpha }(\mathbf{\xi })\partial _{i}\xi
_{\alpha }]\}d^{n+1}x,\qquad d^{n+1}x=\prod\limits_{i=0}^{n}dx^{i},
\label{a2.2}
\end{equation}%
where $j=\{j^{0},\mathbf{j\}}$, $\mathbf{j}=\{j^{\alpha }\}$, $\alpha
=1,2,\ldots n$, $\varphi $, $\mathbf{\xi }=\{\xi _{\alpha }\}$, $\alpha
=1,2,\ldots n$ are functions of $x=\{x^{0},\mathbf{x\}}$, $x^{0}=t$, $%
\mathbf{x}=\{x^{\alpha }\}$, $\alpha =1,2,\ldots n$. A summation is made
over repeating indices: over Latin ones $(0-n)$ and over Greek ones $(1-n)$.
All variables $j=\{j^{0},\mathbf{j\}}$, $\varphi $, $\mathbf{\xi }$ are
functions of $x$. The function $\mathcal{L}(x,j)$ is defined by the relation

\begin{equation}
\mathcal{L}(x,j)=j^{0}L(x^{0},\mathbf{x,j}/j^{0})  \label{a2.3}
\end{equation}%
where $L$ is the Lagrangian of a single system $\mathcal{S}$. The variables $%
j=\{j^{0},\mathbf{j\}}$ describe a flux of particles in the $(n+1)$%
-dimensional space $V$ of coordinates $x$. They are connected with the
variables $\mathbf{x}(t,\mathbf{\xi })$, $d\mathbf{x}(t,\mathbf{\xi })/dt$
by means of relations
\begin{equation}
j^{0}=\det \parallel \xi _{\alpha ,\beta }\parallel ,\qquad \xi _{\beta
,\alpha }\equiv \partial _{\alpha }\xi _{\beta },\qquad \alpha ,\beta
=1,2,\ldots n,\qquad \mathbf{j}=j^{0}{\frac{d\mathbf{x}}{dt}}  \label{a2.4}
\end{equation}%
and the functions $\mathbf{\xi }=\mathbf{\xi }(x)$ are determined as
solutions of the equations
\begin{equation*}
x^{\alpha }=x^{\alpha }(t,\mathbf{\xi }),\qquad \alpha =1,2,\ldots n.
\end{equation*}

Functions $g^{\alpha }(\mathbf{\xi })$ are arbitrary functions of Lagrangian
coordinates $\xi _{\alpha }$, $\alpha =1,2,\ldots n$. Appearance of
arbitrary functions is a result of integration of some dynamic equations
arising in the course of transformation from the action (2.1) to the action
(2.2) (see Appendix A). In turn a possibility of such an integration is
connected with an invariance of dynamic equation with respect to arbitrary
transformation of Lagrangian coordinates $\mathbf{\xi }$

\begin{equation}
\xi _{\alpha }\rightarrow \tilde{\xi}_{\alpha }=\tilde{\xi}_{\alpha }(%
\mathbf{\xi }),\qquad \det \parallel \partial \tilde{\xi}_{\beta }/\partial
\xi _{\alpha }\parallel =1,\qquad \alpha ,\beta =1,2,\ldots  \label{a2.5}
\end{equation}%
Flux $j=\{j^{0},\mathbf{j\}}$ is invariant with respect to the renumbering
transformation (2.5).

A statistical ensemble of classical systems is described conventionally by
the distribution function $F(t,\mathbf{x},\mathbf{p})$ in the phase space of
coordinates $\mathbf{x}$ and momenta $\mathbf{p}$. For the pure statistical
ensemble the distribution function has a special form
\begin{equation}
F(t,\mathbf{x},\mathbf{p})=j^{0}(t,\mathbf{x})\delta \lbrack \mathbf{x}-%
\mathbf{P}(t,\mathbf{x})]  \label{a2.6}
\end{equation}%
where
\begin{equation}
\mathbf{p}=\{p_{\alpha }\},\qquad p_{\alpha }=\partial L/\partial
(dx^{\alpha }/dt),\qquad \alpha =1,2,\ldots n  \label{a2..7}
\end{equation}%
$\mathbf{P}(t,\mathbf{x})=\{P_{\alpha }(t,\mathbf{x})\}$, $\alpha
=1,2,\ldots n$ is a set of functions depending only on $t$ and $\mathbf{x}$,
and $\delta $ denotes the Dirac $\delta $-function. Evolution of this
ensemble is described by the Liouville equation of the form
\begin{equation}
\frac{\partial F}{\partial t}+\frac{\partial H}{\partial p_{\alpha }}\frac{%
\partial F}{\partial x^{\alpha }}-\frac{\partial H}{\partial x^{\alpha }}%
\frac{\partial F}{\partial p_{\alpha }}=0,  \label{a2.8}
\end{equation}
where $H=H(t,\mathbf{x},\mathbf{p})$ is the Hamiltonian function of the
dynamic system $\mathcal{S}$. Dynamic equations for the variables $j^0$, $%
\mathbf{P}(t,\mathbf{x})$ can be obtained by means of a substitution of Eq.
(2.6) into Eq.(2.8).

Among three ways [(2.1),(2.2) and (2.8)] describing the pure statistical
ensemble, the action (2.2) is most convenient for comparison with action of
the system $\mathcal{S}_D$.

Dynamic equations, determined by the action (2.2), have the form
\begin{equation}
{\frac{\delta \mathcal{A}}{\delta \varphi }}=\hbar \partial _{i}j^{i}=0,
\label{a2.9}
\end{equation}%
\begin{equation}
\frac{\delta \mathcal{A}}{\delta j^{i}}=\frac{\partial \mathcal{L}}{\partial
j^{i}}-\hbar \partial _{i}\varphi -\hbar g^{\alpha }(\mathbf{\xi })\xi
_{\alpha ,i}=0,\qquad i=0,1,\ldots n;  \label{a2.10}
\end{equation}%
\begin{equation}
\frac{\delta \mathcal{A}}{\delta \xi _{\alpha }}=-\hbar \left( \frac{%
\partial g^{\beta }(\mathbf{\xi })}{\partial \xi _{\alpha }}-\frac{\partial
g^{\alpha }(\mathbf{\xi })}{\partial \xi _{\beta }}\right) j^{i}\partial
_{i}\xi _{\beta }=0,\qquad \alpha =1,\ldots n;  \label{a2.11}
\end{equation}%
$j^{l}$, $l=0,1,\ldots n$ is the current $(n+1)$-vector in the space $V$.
The vector field $j^{i}$ is tangent to the trajectories of systems in $V$.
According to Eq.(2.3) the vector

\begin{equation*}
p_{l}=\frac{\partial \mathcal{L}}{\partial j^{l}},\qquad l=0,1,\ldots
n;\qquad p_{0}=\left[ L-\frac{\partial L}{\partial (dx^{\alpha }/dt)}\right]
_{dx^{\alpha }/dt=j^{\alpha }/j^{0}};
\end{equation*}%
\begin{equation}
p_{\alpha }=\left[ \frac{\partial L}{\partial (dx^{\alpha }/dt)}\right]
_{dx^{\alpha }/dt=j^{\alpha }/j^{0}},\qquad \alpha =1,2,\ldots n
\label{a2.12}
\end{equation}%
associates with the canonical momentum of a single system of the ensemble.
Thus, according to Eq.(2.10)
\begin{equation}
p_{i}=\hbar \left( \partial _{i}\varphi +g^{\alpha }(\xi )\mathbf{\xi }%
_{\alpha ,i}\right) ,\qquad i=0,1,2,\ldots n  \label{a2.13}
\end{equation}%
where $\hbar $ is a constant having a dimensionality of the action. In this
case $\varphi $, $g^{\alpha }$, and $\mathbf{\xi }$ can be considered as
dimensionless quantities. $\hbar $ can be treated as the Planck constant,
although in the given case it has no quantum meaning.

Eliminating variables $\varphi $ and $\mathbf{\xi }$ from Eqs. (2.10),
(2.11), one obtains the equations
\begin{equation}
j^{i}[\partial _{i}p_{k}-\partial _{k}p_{i}]=j^{i}[\partial _{i}\frac{%
\partial \mathcal{L}(x,j)}{\partial j^{k}}-\partial _{k}\frac{\partial
\mathcal{L}(x,j)}{\partial j^{i}}]=0,\qquad k=0,1,\ldots n  \label{a2.14}
\end{equation}%
Together with Eq.(2.9) the equations (2.14) form a system of $n+1$ dynamic
equations for some kind of a fluid described by the current vector $j^{i}$.

At the linear transformation of coordinates $x^{i}$, $i=0,1,\ldots n$ the
current vector $j^{l}$, $l=0,1,\ldots n$ transforms as a vector. In this
case the Lagrangian coordinates transform as scalars. But the consideration
of $\mathbf{\xi }$ as scalars is rather conventional, because of the
renumbering transformation (2.5).

At the transformation (2.5) one obtains

\begin{equation}
j^{l}g^{\alpha }(\mathbf{\xi })\partial _{l}\xi _{\alpha }\rightarrow j^{l}%
\tilde{g}^{\alpha }(\mathbf{\tilde{\xi}})\partial _{l}\tilde{\xi}_{\alpha
},\qquad \tilde{g}^{\alpha }(\tilde{\mathbf{\xi }})=g^{\beta }(\mathbf{\xi })%
{\frac{\partial \xi _{\beta }}{\partial \tilde{\xi}_{\alpha }}}
\label{a2.15}
\end{equation}%
In other words, $g^{\alpha }(\mathbf{\xi })$ transform at the renumbering
transformation as components of a vector in the space of $\mathbf{\xi }$.
Combining any linear transformation of coordinates $x^{i}$ with some
renumbering transformation, one can ascribe practically arbitrary
transformation properties to the Lagrangian coordinates $\mathbf{\xi }$.

Any renumbering transformation (2.5) is a kind of a gauge transformation,
because the renumbering changes a description of the state of the
statistical ensemble without changing the state itself. It is easy to verify
that a set of the renumbering transformations forms a group.

For the pure statistical ensemble of classical pointlike charged particles
moving in the given electromagnetic field the action (2.2) takes the form

\begin{equation}
\mathcal{A}_{cl}[j,\xi ,\varphi ]=\int [-m\sqrt{j^{l}j_{l}}%
-eA_{l}j^{l}-\hbar j^{i}(\partial _{i}\varphi +g^{\alpha }(\mathbf{\xi })\xi
_{\alpha ,i})]d^{4}x  \label{a2.16}
\end{equation}%
where the speed of the light $c=1,$ and summation is made over repeated
indices (0 - 3) for the Latin indices and (1 - 3) for the Greek ones.

\section{Transformation of variables}

Transforming the Dirac equation (1.1) to the new variables (1.2), one uses
the action for the equation (1.1)
\begin{equation}
\mathcal{A}_{D}[\bar{\psi},\psi ]=\int (-m\bar{\psi}\psi +{\frac{i}{2}}\hbar
\bar{\psi}\gamma ^{l}\partial _{l}\psi -{\frac{i}{2}}\hbar \partial _{l}\bar{%
\psi}\gamma ^{l}\psi -eA_{l}\bar{\psi}\gamma ^{l}\psi )d^{4}x  \label{a3.1}
\end{equation}%
Expressing the variables (1.2) through the wave function $\psi $,
one uses the technique, where the wave function is considered as a
Clifford number
with 16 base units: $I,$ $\gamma ^{i}$, $\gamma ^{ik}$, $\gamma ^{ikl}$, $%
\gamma ^{iklm}$ (all indices are different, and Clifford numbers satisfy (%
\ref{a1.3})). Reduction of the Clifford numbers is realized by means of zero
divisors [4,5].

Let us introduce matrices $\gamma _{5}$, $\mathbf{\sigma }=\{\sigma _{\alpha
}\},$ $\alpha =1,2,3$
\begin{equation*}
\gamma _{5}=\gamma ^{0123}\equiv \gamma ^{0}\gamma ^{1}\gamma ^{2}\gamma
^{3},\qquad \sigma _{1}=-i\gamma ^{23},\qquad \sigma _{2}=-i\gamma
^{31},\qquad \sigma _{3}=-i\gamma ^{12},
\end{equation*}%
\begin{equation}
\gamma _{5}\sigma _{\alpha }=\sigma _{\alpha }\gamma _{5},\qquad \gamma
^{0\alpha }=-i\gamma _{5}\sigma _{\alpha },\qquad \alpha =1,2,3;
\label{a3.2}
\end{equation}%
\begin{equation*}
\gamma ^{0}\mathbf{\sigma }=\mathbf{\sigma }\gamma ^{0},\qquad \gamma
^{0}\gamma _{5}=-\gamma _{5}\gamma ^{0}
\end{equation*}%
According to Eqs. (1.3), (3.2) the matrices $\mathbf{\sigma }=\{\sigma
_{\alpha }\}$, $\alpha =1,2,3$ satisfy the relation
\begin{equation}
\sigma _{\alpha }\sigma _{\beta }=\delta _{\alpha \beta }+i\varepsilon
_{\alpha \beta \gamma }\sigma _{\gamma },\qquad \alpha ,\beta =1,2,3
\label{a3.3}
\end{equation}%
where $\varepsilon _{\alpha \beta \gamma }$ is the antisymmetric
pseudo-tensor of Levi-Chivita $(\varepsilon _{123}=1)$.

Let us define the wave function $\psi $ in the form

\begin{equation*}
\psi =Ae^{i\varphi +{\frac{1}{2}}\gamma _{5}\kappa }e^{-{\frac{i}{2}}\gamma
_{5}\mathbf{\sigma \eta }}e^{{\frac{i}{2}}\mathbf{\sigma \zeta }}\Pi
\end{equation*}%
\begin{equation}
\bar{\psi}=\Pi \psi ^{\ast }\gamma ^{0},\qquad \psi ^{\ast }=A\Pi e^{-{\frac{%
i}{2}}\mathbf{\sigma \zeta }}e^{-{\frac{i}{2}}\gamma _{5}\mathbf{\sigma \eta
}}e^{-i\varphi -{\frac{1}{2}}\gamma _{5}\kappa }  \label{a3.4}
\end{equation}%
where (*) means the Hermitian conjugation, and
\begin{equation}
\Pi ={\frac{1}{4}}(1+\gamma ^{0})(1+\mathbf{z\sigma }),\qquad \mathbf{z}%
=\{z^{\alpha }\}=\text{const},\qquad \alpha =1,2,3;\qquad
\mathbf{z}^{2}=1 \label{a3.5}
\end{equation}%
is a zero divisor. The quantities $A$, $\kappa $, $\varphi $, $\mathbf{\eta }%
=\{\eta ^{\alpha }\}$, $\mathbf{\zeta }=\{\zeta ^{\alpha }\}$, $\alpha =1,2,3
$ are nine real parameters, determining the wave function $\psi $.

Using relations (3.2), (3.3), (3.5), it is easy to verify that
\begin{equation*}
\Pi ^{2}=\Pi ,\qquad \gamma _{0}\Pi =\Pi ,\qquad \mathbf{z\sigma }\Pi =\Pi
,\qquad \Pi \gamma _{5}\Pi =0,
\end{equation*}%
\begin{equation}
\Pi \sigma _{\alpha }\Pi =z^{\alpha }\Pi ,\qquad \alpha =1,2,3.  \label{a3.6}
\end{equation}%
Generally, the wave functions $\psi ,\psi ^{\ast }$ defined by Eq.(3.4) are $%
4\times 4$ complex matrices. In the proper representation, where $\Pi $ has
the form
\begin{equation}
\Pi =\left(
\begin{array}{cccc}
1 & 0 & 0 & 0 \\
0 & 0 & 0 & 0 \\
0 & 0 & 0 & 0 \\
0 & 0 & 0 & 0%
\end{array}%
\right)  \label{a3.7}
\end{equation}%
the $\psi ,\psi ^{\ast }$ have the form

\begin{equation}
\psi =\left(
\begin{array}{cccc}
\psi _{1} & 0 & 0 & 0 \\
\psi _{2} & 0 & 0 & 0 \\
\psi _{3} & 0 & 0 & 0 \\
\psi _{4} & 0 & 0 & 0%
\end{array}%
\right) ,\qquad \psi ^{\ast }=\left(
\begin{array}{cccc}
\psi _{1}^{\ast } & \psi _{2}^{\ast } & \psi _{3}^{\ast } & \psi _{4}^{\ast }
\\
0 & 0 & 0 & 0 \\
0 & 0 & 0 & 0 \\
0 & 0 & 0 & 0%
\end{array}%
\right)  \label{a3.8}
\end{equation}%
Their product $\psi ^{\ast }O\psi $ has the form
\begin{equation}
\psi ^{\ast }O\psi =\left(
\begin{array}{cccc}
a & 0 & 0 & 0 \\
0 & 0 & 0 & 0 \\
0 & 0 & 0 & 0 \\
0 & 0 & 0 & 0%
\end{array}%
\right) =a\Pi =\Pi a  \label{a3.9}
\end{equation}%
where $O$ is an arbitrary $4\times 4$ matrix and $a$ is a complex quantity.
If $f$ is an analytical function having the property $f(0)=0,$ then the
function $f(a\Pi )$ of a $4\times 4$ matrix of the type (3.9) is a matrix $%
f(a)\Pi $ of the same type. For this reason one will not distinguish between
the complex quantity $a$ and the complex $4\times 4$ matrix $a\Pi $. In the
final expressions of the type $a\Pi $ ($a$ is a complex quantity) the
multiplier $\Pi $ will be omitted.

By means of relations (3.2), (3.6) one can reduce any Clifford number $\Pi
O\Pi $ to the form (3.9), without using any concrete form of the $\gamma $%
-matrix representation. This property will be used in our calculations.

By means of (\ref{a3.4}) the variables $\bar{\psi}\psi $, $j^{l}$, $S^{l}$, $%
l=0,1,2,3$ defined by expressions (1.2) can be presented in the form
\begin{equation}
\bar{\psi}\psi =\psi ^{\ast }\gamma ^{0}\psi =A^{2}\Pi e^{\gamma _{5}\kappa
}\Pi =A^{2}\cos \kappa \Pi  \label{a3.10}
\end{equation}%
\begin{equation}
j^{0}\Pi =\bar{\psi}\gamma ^{0}\psi =A^{2}\Pi e^{-{\frac{i}{2}}\mathbf{%
\sigma \zeta }}e^{-i\gamma _{5}\mathbf{\sigma \eta }}e^{{\frac{i}{2}}\mathbf{%
\sigma \zeta }}\Pi =A^{2}\Pi e^{-i\gamma _{5}\mathbf{\Sigma \eta }}\Pi
=A^{2}\cosh (\eta )\Pi  \label{a3.11}
\end{equation}%
where
\begin{equation}
\mathbf{\Sigma }=\{\Sigma _{1},\Sigma _{2},\Sigma _{3}\},\qquad \Sigma
_{\alpha }=e^{-{\frac{i}{2}}\mathbf{\sigma \zeta }}\sigma _{\alpha }e_{,}^{{%
\frac{i}{2}}\mathbf{\sigma \zeta }}\qquad \alpha =1,2,3;  \label{a3.12}
\end{equation}%
\begin{equation*}
\eta =\sqrt{\mathbf{\eta }^{2}}=\sqrt{\eta ^{\alpha }\eta ^{\alpha }}
\end{equation*}%
The matrix $\Sigma _{\alpha }$ satisfy the same relations (3.3), as $\sigma
_{\alpha }$ do.

In the same way one obtains
\begin{equation}
\begin{array}{ll}
j^{\alpha }\Pi & =\psi ^{\ast }\gamma ^{0\alpha }\psi \Pi =A^{2}\Pi e^{-{%
\frac{i}{2}}\gamma _{5}\mathbf{\Sigma \eta }}(-i\gamma _{5}\Sigma _{\alpha
})e^{-{\frac{i}{2}}\gamma _{5}\mathbf{\Sigma \eta }}\Pi = \\
& =A^{2}\Pi (\cosh {\frac{\eta }{2}}-i\gamma _{5}\mathbf{\Sigma }\mathbf{v}%
\sinh {\frac{\eta }{2}})(-i\gamma _{5}\Sigma _{\alpha })(\cosh {\frac{\eta }{%
2}}-i\gamma _{5}\mathbf{\Sigma }\mathbf{v}\sinh {\frac{\eta }{2}})\Pi = \\
& =A^{2}\sinh (\eta )v^{\alpha }\Pi ,\qquad \alpha =1,2,3,%
\end{array}
\label{a3.13}
\end{equation}%
where
\begin{equation}
\mathbf{v}=\{v^{\alpha }\},\qquad v^{\alpha }=\eta ^{\alpha }/\eta ,\qquad
\alpha =1,2,3;\qquad \mathbf{v}^{2}=1.  \label{a3.14}
\end{equation}

\begin{equation*}
\begin{array}{lll}
S^{0}\Pi & = & \psi ^{\ast }(-i\gamma _{5})\psi =A^{2}\Pi (-i\gamma
_{5})e^{-i\gamma _{5}\mathbf{\Sigma \eta }}\Pi = \\
& = & A^{2}\Pi \sinh (\eta )\mathbf{\Sigma }\mathbf{v}\Pi =A^{2}\sinh (\eta )%
\mathbf{\xi }\Pi ,%
\end{array}%
\end{equation*}%
\begin{equation}
\begin{array}{ll}
S^{\alpha }\Pi = & \psi ^{\ast }\sigma _{\alpha }\psi \Pi =A^{2}\Pi e^{{%
\frac{i}{2}}\gamma _{5}\mathbf{\Sigma \eta }}\Sigma _{\alpha }e^{{\frac{i}{2}%
}\gamma _{5}\mathbf{\Sigma \eta }}\Pi = \\
& A^{2}[\xi ^{\alpha }+(\cosh \eta -1)v^{\alpha }(\mathbf{v\xi })]\Pi
,\qquad \alpha =1,2,3.%
\end{array}
\label{a3.15}
\end{equation}%
Here $\mathbf{\xi }=\{\xi ^{\alpha }\}$, $\alpha =1,2,3$, are determined by
the relation
\begin{equation}
\xi ^{\alpha }\Pi =\Pi \Sigma _{\alpha }\Pi ,\qquad \alpha =1,2,3
\label{a3.16}
\end{equation}%
It follows from Eqs.(3.11), (3.13)
\begin{equation}
j^{i}j_{i}\Pi =A^{4}\Pi ,\qquad A=(j^{l}j_{l})^{1/4}\equiv \rho ^{1/2}
\label{a3.17}
\end{equation}%
According to Eqs.(3.12), (3.16) one obtains

\begin{equation}
\mathbf{\xi }\Pi =\{[\mathbf{z}-\mathbf{n}(\mathbf{nz})]\cos \zeta +(\mathbf{%
z\times n})\sin \zeta +\mathbf{n}(\mathbf{nz})\}\Pi  \label{a3.18}
\end{equation}%
where
\begin{equation*}
\mathbf{\xi }=\{\xi ^{\alpha }\},\qquad \alpha =1,2,3;\qquad \zeta =\sqrt{%
\mathbf{\zeta }^{2}}=\sqrt{\zeta ^{\alpha }\zeta ^{\alpha }},
\end{equation*}%
\begin{equation}
\mathbf{n}=\mathbf{\zeta }/\zeta ,\qquad \mathbf{n}^{2}=1  \label{a3.19}
\end{equation}%
The wave function (3.4) depends on 9 real parameters: $A$, $\varphi $, $%
\kappa $, $\eta _{\alpha }$, $\zeta _{\alpha }$, $\alpha =1,2,3$. The wave
function has 8 real independent components, and not all parameters $%
A,\varphi ,\kappa ,\eta ^{\alpha },\zeta ^{\alpha }$, $\alpha =1,2,3$ are
independent. Let us fix one of the parameters, namely let us set
\begin{equation}
\zeta =\pi  \label{a3.20}
\end{equation}%
Then Eq.(3.18) takes the form
\begin{equation}
\mathbf{\xi }=2\mathbf{n}(\mathbf{nz})-\mathbf{z}  \label{a3.21}
\end{equation}%
It can be solved with respect to $\mathbf{n}=\mathbf{\zeta }/\pi $. One
obtains

\begin{equation}
\mathbf{\zeta }/\pi =\mathbf{n}=(\mathbf{\xi }+\mathbf{z})[2(1+\mathbf{z\xi }%
)]^{-1/2}  \label{a3.22}
\end{equation}%
Using Eqs.(3.11), (3.13), (3.17), (3.22), one can express parameters $%
A,\varphi ,\kappa ,\eta ^{\alpha },\zeta ^{\alpha }$, $\alpha =1,2,3$
describing the wave function, through the variables $j^{i},S^{i},\varphi
,\kappa $. One obtains
\begin{equation}
\xi ^{\alpha }=\rho ^{-1}[S^{\alpha }-\frac{j^{\alpha }(S^{\beta }j^{\beta })%
}{j^{0}(j^{0}+\rho )}],\qquad \alpha =1,2,3;\qquad \rho \equiv \sqrt{%
j^{l}j_{l}}  \label{a3.23}
\end{equation}%
\begin{equation}
\cosh \eta =j^{0}/\rho ,\qquad v^{\alpha }=\frac{\eta ^{\alpha }}{\eta }=%
\frac{j^{\alpha }}{\sqrt{(j^{0})^{2}-j^{l}j_{l}}},\qquad \alpha =1,2,3
\label{a3.24}
\end{equation}%
Using Eqs.(3.4), (3.20) - (3.24), one can present the wave function (3.4) in
terms of variables $j^{i},S^{i},\kappa ,\varphi $:
\begin{equation}
\psi =\frac{ie^{i\varphi +{\frac{1}{2}}\gamma _{5}\kappa }}{2\sqrt{(1+%
\mathbf{\xi })(j^{0}+\rho )}}[(j^{0}+\rho )(1+\mathbf{\xi \sigma })-i\gamma
_{5}S^{0}-i\gamma _{5}(\mathbf{j}+i\rho ^{-1}\mathbf{j\times S})\mathbf{%
\sigma }]\Pi  \label{a3.25}
\end{equation}%
where $\mathbf{\xi }$ is expressed through $j^{i},S^{i}$ by means of
Eq.(3.23), the symbol $\times $ means a vector product, and $\Pi $ is
defined by Eq.(3.5).

Any expression of the form $\bar{\psi}O\psi $, where $O$ is an arbitrary
combination of $\gamma $-matrices, can be expressed through the variables $%
S^{i},j^{i},\kappa $. For instance,
\begin{equation}
{\frac{i}{2}}\bar{\psi}(\gamma ^{l}\gamma ^{k}-\gamma ^{k}\gamma ^{l})\psi ={%
\frac{1}{2\rho }}[(j^{k}S^{l}-j^{l}S^{k})\sin \kappa +\varepsilon
^{lkim}j_{i}S_{m}\cos \kappa ]\Pi .  \label{a3.26}
\end{equation}

\section{Transformation of the action}

Now let us calculate the expression (two middle terms of Eq.(3.1))
\begin{equation}
\begin{array}{ll}
{\frac{i}{2}}\hbar \bar{\psi}\gamma ^{l}\partial _{l}\psi +\text{h.c.}= & {%
\frac{i}{2}}\hbar \psi ^{\ast }\left[ \left( \partial _{0}-i\gamma _{5}%
\mathbf{\sigma \nabla }\right) \left( i\varphi +\frac{1}{2}\gamma _{5}\kappa
\right) \right] \psi +\text{h.c.} \\
& +\frac{i}{2}\hbar A^{2}\Pi e^{-\frac{i}{2}\mathbf{\sigma \zeta }}e^{-\frac{%
i}{2}\gamma _{5}\mathbf{\sigma \eta }}(\partial _{0}-i\gamma _{5}\mathbf{%
\sigma }\nabla )(e^{-\frac{i}{2}\gamma _{5}\mathbf{\sigma \eta }}e^{{\frac{i%
}{2}}\mathbf{\sigma \zeta }})\Pi +\text{h.c.}%
\end{array}
\label{a4.1}
\end{equation}%
where \textquotedblright h.c.\textquotedblright\ means the term obtained
from the previous one by the Hermitian conjugation. Using relations
(3.11)-(3.15), the expression (4.1) reduces to the form
\begin{equation}
\begin{array}{ll}
{\frac{i}{2}}\hbar \bar{\psi}\gamma ^{l}\partial _{l}\psi +\text{h.c.}= &
-\hbar j^{l}\partial _{l}\varphi -{\frac{1}{2}}\hbar S^{l}\partial _{l}\kappa
\\
& +{\frac{i}{2}}\hbar A^{2}\Pi e^{-\frac{i}{2}\gamma _{5}\mathbf{\Sigma \eta
}}(\partial _{0}-i\gamma _{5}\mathbf{\Sigma }\nabla )e^{-\frac{i}{2}\gamma
_{5}\mathbf{\Sigma \eta }}\Pi +\text{h.c.} \\
& +{\frac{i}{2}}\hbar A^{2}\Pi e^{-i\gamma _{5}\mathbf{\Sigma \eta }}e^{-%
\frac{i}{2}\mathbf{\sigma \zeta }}\partial _{0}e^{\frac{i}{2}\mathbf{\sigma
\zeta }}\Pi +\text{h.c.} \\
& +{\frac{i}{2}}\hbar A^{2}\Pi e^{-\frac{i}{2}\gamma _{5}\mathbf{\Sigma \eta
}}(-i\gamma _{5}\Sigma _{\alpha })e^{-\frac{i}{2}\gamma _{5}\mathbf{\Sigma
\eta }}e^{-\frac{i}{2}\mathbf{\sigma \zeta }}\partial _{\alpha }e^{\frac{i}{2%
}\mathbf{\sigma \zeta }}\Pi +\text{h.c.}%
\end{array}
\label{a4.2}
\end{equation}%
where the matrix $\mathbf{\Sigma }$ is not differentiated.

Taking into account the fourth Eq.(3.6) and Eqs. (3.11), (3.13), the
expression (4.2) reduces to the form
\begin{equation}
\begin{array}{ll}
{\frac{i}{2}}\hbar \bar{\psi}\gamma ^{l}\partial _{l}\psi +\text{h.c.}= &
-\hbar j^{l}\partial _{l}\varphi -{\frac{1}{2}}\hbar S^{l}\partial
_{l}\kappa +{\frac{i}{2}}\hbar j^{l}\Pi e^{-{\frac{i}{2}}\mathbf{\sigma
\zeta }}\partial _{l}e^{{\frac{i}{2}}\mathbf{\sigma \zeta }}\Pi +\text{h.c.}
\\
& +{\frac{i}{2}}\hbar A^{2}\Pi e^{-{\frac{i}{2}}\gamma _{5}\mathbf{\Sigma
\eta }}(\partial _{0}-i\gamma _{5}\mathbf{\Sigma }\nabla )e^{-{\frac{i}{2}}%
\gamma _{5}\mathbf{\Sigma \eta }}\Pi +\text{h.c.}%
\end{array}
\label{a4.3}
\end{equation}%
Substituting the relation (3.20) into the third term of Eq.(4.3), one
obtains by means of Eq.(3.22)
\begin{eqnarray}
{\frac{i}{2}}\hbar j^{l}\Pi e^{-\frac{i}{2}\mathbf{\sigma \zeta }}\partial
_{l}e^{\frac{i}{2}\mathbf{\sigma \zeta }}\Pi +\text{h.c.} &=&{\frac{i}{2}}%
\hbar j^{l}\Pi \sigma _{\alpha }\sigma _{\beta }n_{\alpha }\partial
_{l}n_{\beta }\Pi +\text{h.c.}  \notag \\
&=&-{\frac{\hbar j^{l}}{{2(1+\mathbf{\xi z})}}}\varepsilon _{\alpha \beta
\gamma }\xi ^{\alpha }\partial _{l}\xi ^{\beta }z^{\gamma }\Pi =-\hbar
j^{l}g_{\alpha }(\mathbf{\xi })\partial _{l}\xi ^{\alpha }\Pi  \label{a4.4}
\end{eqnarray}%
where
\begin{equation}
g_{\alpha }(\mathbf{\xi })=-{\frac{1}{2}}\varepsilon _{\alpha \beta \gamma
}\xi ^{\beta }z^{\gamma }(1+\mathbf{\xi z})^{-1}  \label{a4.5}
\end{equation}%
Calculation of the last term of Eq.(4.3) leads to the following result
\begin{equation}
\begin{array}{ll}
F_{4}= & \frac{i}{2}\hbar A^{2}\Pi e^{-\frac{i}{2}\gamma _{5}\mathbf{\Sigma
\eta }}(\partial _{0}-i\gamma _{5}\mathbf{\Sigma }\nabla )e^{-\frac{i}{2}%
\gamma _{5}\mathbf{\Sigma \eta }}\Pi +\text{h.c.}= \\
& -{\frac{1}{2}}\hbar A^{2}\varepsilon _{\alpha \beta \gamma }[\partial
_{\alpha }\eta v^{\beta }\xi ^{\gamma }+\sinh \eta \partial _{\alpha
}v^{\beta }\xi ^{\gamma }+2\sinh ^{2}({\frac{\eta }{2}})v^{\alpha }\partial
_{0}v^{\beta }\xi ^{\gamma }]\Pi%
\end{array}
\label{a4.6}
\end{equation}%
Let us introduce two constant vectors
\begin{equation}
f^{i}=\{1,0,0,0,\},\qquad z^{i}=\{0,z^{1},z^{2},z^{3}\}  \label{a4.7}
\end{equation}%
which satisfy the following conditions
\begin{equation}
f^{l}f_{l}=1,\qquad f^{l}z_{l}=0,\qquad z^{l}z_{l}=-1.  \label{a4.8}
\end{equation}%
By means of relations (3.11), (3.13) and (4.7) the expression (4.6) reduces
to the form
\begin{equation}
F_{4}=-\frac{\hbar }{2(\rho +f^{s}j_{s})}\varepsilon _{iklm}[\partial
^{k}(j^{i}+f^{i}\rho )](j^{l}+f^{l}\rho )[\xi ^{m}-f^{m}(\xi ^{s}f_{s})]
\label{a4.9}
\end{equation}%
where $\varepsilon _{iklm}$ is the Levi-Chivita pseudo-tensor $(\varepsilon
_{0123}=1)$ and $\xi ^{m}=\{\xi ^{0},\mathbf{\xi }\}$. The value of $\xi
^{0} $ is unessential.

Let us introduce the unit timelike vector
\begin{equation}
q^{i}\equiv \frac{j^{i}+f^{i}\rho }{\sqrt{(j^{l}+f^{l}\rho )(j_{l}+f_{l}\rho
)}}=\frac{j^{i}+f^{i}\rho }{\sqrt{2\rho (\rho +j^{l}f_{l})}}  \label{a4.10}
\end{equation}%
and two spacelike vectors
\begin{equation}
\nu ^{i}=\xi ^{i}-f^{i}(\xi ^{s}f_{s}),\qquad i=0,1,2,3;\qquad \nu ^{i}\nu
_{i}=-1,  \label{a4.11}
\end{equation}%
\begin{equation}
\mu ^{i}\equiv \frac{\nu ^{i}}{\sqrt{-(\nu ^{l}+z^{l})(\nu _{l}+z_{l})}}=%
\frac{\nu ^{i}}{\sqrt{2(1-\nu ^{l}z_{l})}}={\frac{\nu ^{i}}{\sqrt{2(1+%
\mathbf{\xi z})}}}.  \label{a4.12}
\end{equation}%
Then according to Eqs.(4.3), (4.4), (4.9) the relation (4.2) can be
presented in the form
\begin{equation}
{\frac{i}{2}}\hbar \bar{\psi}\gamma ^{l}\partial _{l}\psi +\text{h.c.}%
=-\hbar j^{l}[\partial _{l}\varphi +g_{\alpha }(\mathbf{\xi })\partial
_{l}\xi ^{\alpha }]-{\frac{\hbar }{2}}S^{l}\partial _{l}\kappa +\hbar \rho
\varepsilon _{iklm}q^{i}(\partial ^{k}q^{l})\nu ^{m}  \label{a4.13}
\end{equation}%
where the second term can be written also in the covariant form
\begin{equation}
-\hbar j^{l}g_{\alpha }(\xi )\partial _{l}\xi ^{\alpha }=-\hbar
j^{l}\varepsilon _{jksm}\mu ^{j}(\partial _{l}\mu ^{k})f^{s}z^{m}
\label{a4.14}
\end{equation}

Now by means of relations (3.10)-(3.12) one can present the action (3.1) in
the form
\begin{equation}
\mathcal{A}_{D}[j,\varphi ,\kappa ,\mathbf{\xi }]=\int (\mathcal{L}_{cl}+%
\mathcal{L}_{q1}+\mathcal{L}_{q2})d^{4}x  \label{a4.15}
\end{equation}%
\begin{equation}
\mathcal{L}_{cl}=-m\rho -eA_{l}j^{l}-\hbar j^{i}[\partial _{i}\varphi
+g_{\alpha }(\mathbf{\xi })\partial _{i}\xi ^{\alpha }],\qquad \rho \equiv
\sqrt{j^{l}j_{l}}  \label{a4.16}
\end{equation}%
\begin{equation}
\mathcal{L}_{q1}=2m\rho \sin ^{2}({\frac{\kappa }{2}})-{\frac{\hbar }{2}}%
S^{l}\partial _{l}\kappa ,  \label{a4.17}
\end{equation}%
\begin{equation}
\mathcal{L}_{q2}=\hbar \rho \varepsilon _{iklm}q^{i}(\partial ^{k}q^{l})\nu
^{m}  \label{a4.18}
\end{equation}%
$g_{\alpha }(\mathbf{\xi })$ is defined by the equation (4.5). $S^{l}$, $%
l=0,1,2,3$ are considered as functions of $j^{l}$ and $\mathbf{\xi }$,
defined by the relations
\begin{equation}
S^{0}=\mathbf{j\xi },\qquad S^{\alpha }=\rho \xi ^{\alpha }+\frac{(\mathbf{%
j\xi })j^{\alpha }}{\rho +j^{k}f_{k}},\qquad \alpha =1,2,3  \label{a4.19}
\end{equation}%
obtained from Eqs (1.4), (3.23). Not all variables $\mathbf{\xi }=\{\xi
^{\alpha }\}$, $\alpha =1,2,3$ are independent, because they satisfy the
restriction
\begin{equation}
\mathbf{\xi }^{2}=\xi ^{\alpha }\xi ^{\alpha }=1  \label{a4.20}
\end{equation}%
as it follows from Eqs.(3.5), (3.19), (3.21). Variation of the action (4.15)
with respect to $\xi ^{\alpha }$, $\alpha =1,2,3$ under the condition (4.20)
leads to the dynamic equations

\begin{equation}
{\frac{\delta \mathcal{A}_{D}}{\delta \xi ^{\alpha }}}(\delta ^{\alpha \beta
}-\xi ^{\alpha }\xi ^{\beta })=0,\qquad \beta =1,2,3  \label{a4.21}
\end{equation}%
There are only two independent equations among the equations (4.21), because
a contraction of Eq.(4.21) with $\xi ^{\beta }$ leads to an identity. Note
that the term (4.16) of the action (4.15) coincides with the Lagrangian of
the action (2.16) for the statistical ensemble of classical pointlike
particles.

Eliminating $\kappa $ from the action (4.15) and the dynamic equation $%
\delta \mathcal{A}_{D}/\delta \kappa =0$, one can write the action (4.15) in
the form
\begin{equation}
\mathcal{A}_{D}[j,\varphi ,\mathbf{\xi }]=\int (\mathcal{L}_{cl}^{\prime
\prime }+\mathcal{L}_{q1}^{\prime \prime }+\mathcal{L}_{q2})d^{4}x
\label{a4.22}
\end{equation}%
\begin{equation*}
\mathcal{L}_{cl}^{\prime \prime }+\mathcal{L}_{q1}^{\prime \prime }=-\sqrt{%
m^{2}j^{i}j_{i}-{\frac{\hbar ^{2}}{4}}(\partial _{i}S^{i})^{2}}-{\frac{\hbar
}{2}}\partial _{i}S^{i}\arcsin {(}\frac{\hbar \partial _{i}S^{i}}{2m\rho }{)}%
-
\end{equation*}%
\begin{equation}
-ej^{i}A_{i}-\hbar j^{i}[\partial _{i}\varphi +g_{\alpha }(\mathbf{\xi }%
)\partial _{i}\xi ^{\alpha }]  \label{a4.23}
\end{equation}%
$S^{i}$, $i=0,1,2,3$ are functions of $\mathbf{\xi }$ and $j^{i}$ which are
determined by Eqs.(4.19). $\mathcal{L}_{q2}$ is determined by Eq. (4.18).

\section{Classical Part of the Action}

The classical part of the action $\mathcal{A}_D$ can be separated out either
by vanishing $\hbar$, or by a usage of the quasi-uniform state of the
system, when all spatial gradients are small, and quantum effects disappear.

In the non-relativistic case a quasi-uniform state satisfies the condition
\begin{equation}
\mid \frac{\hbar }{m}\nabla u\mid \ll \mid u\mid ,\qquad
u=j^{0},j^{1},j^{2},j^{3},\kappa .  \label{a5.1}
\end{equation}%
which is written in the coordinate system, where $\mid \mathbf{j}\mid \ll
\mid j^{0}\mid $. In the general case such a coordinate system does not
exist, and the condition (5.1) is written in the form
\begin{equation}
\mid \frac{\hbar }{m}l^{i}\partial _{i}u\mid \ll \mid u\mid ,\qquad
u=j^{0},j^{1},j^{2},j^{3},\kappa ,  \label{a5.2}
\end{equation}%
where $l^{i}$ is any unit vector orthogonal to $j^{i}$
\begin{equation}
l^{i}j_{i}=0,\qquad l^{i}l_{i}=-1.  \label{a5.3}
\end{equation}%
It means that all derivatives across the direction of the vector $j^{i}$ are
small.

Let us represent all derivatives in the Eq.(4.18) in the form
\begin{equation}
\partial ^{k}q^{l}=\partial _{\perp }^{k}q^{l}+\frac{j^{k}j^{s}}{\rho ^{2}}%
\partial _{s}q^{l},\qquad \partial _{\perp }^{k}\equiv \partial ^{k}-\frac{%
j^{k}j^{s}}{\rho ^{2}}\partial _{s}  \label{a5.4}
\end{equation}%
The first term in rhs of the first equation (5.4) describes a transversal
part of the derivative, i.e. a derivative in the direction orthogonal to the
vector $j^{i}$.

According to Eq.(5.2) all terms containing the transversal derivative $%
\partial _{\perp }^{k}$ are small with respect to the first term of
Eq.(4.16). Indeed, an estimation of the transversal part of the Lagrangian
(4.18) has the form
\begin{equation}
\mid \hbar \rho \varepsilon _{iklm}q^{i}(\partial _{\perp }^{k}q^{l})\nu
^{m}\mid \cong \hbar \rho \mid \sum_{s=0}^{3}l_{(s)}^{i}\partial
_{i}q^{s}\mid \ll m\rho ,  \label{a5.5}
\end{equation}%
where $l_{(s)}^{i},s=0,1,2,3$ are unit vectors orthogonal to $j^{i}$

\begin{equation}
l_{(s)}^{l}=\frac{\varepsilon _{i.sm}^{.j}q^{i}\nu ^{m}(\delta _{j}^{l}-\rho
^{-2}j_{j}j^{l})}{\mid \varepsilon _{ijsm}q^{i}\nu ^{m}(g^{l^{\prime
}j}-\rho ^{-2}j^{j}j^{l^{\prime }})\varepsilon _{i^{\prime }l^{\prime
}sm^{\prime }}q^{i^{\prime }}\nu ^{m^{\prime }}\mid ^{1/2}},\qquad s=0,1,2,3
\label{a5.6}
\end{equation}%
\centerline{(no summation over $s$)}

Neglecting the transversal part and using Eqs.(4.10), (4.11), one can write
Eq.(4.18) in the form
\begin{equation*}
\mathcal{L}_{qu2}=\hbar \rho ^{-1}j^{s}\varepsilon
_{iklm}q^{i}j^{k}(\partial _{s}q^{l})\nu ^{m}={\frac{\hbar j^{s}}{\sqrt{%
2\rho (\rho +j^{j}f_{j})}}}\varepsilon _{iklm}f^{i}j^{k}\partial
_{s}q^{l}\xi ^{m}
\end{equation*}%
\begin{equation}
=\frac{\hbar j^{i}}{{2\rho (\rho +j^{j}f_{j})}}\varepsilon
_{klsm}j^{k}\partial _{i}j^{l}f^{s}\xi ^{m}  \label{a5.7}
\end{equation}

The relation (4.17) can be written in the form
\begin{equation}
\mathcal{L}_{q1}=m\rho (2\sin ^{2}{\frac{\kappa }{2}}-\frac{\hbar }{2m}%
w^{i}\partial _{i}\kappa ),\qquad w^{i}=\frac{S^{i}}{\sqrt{-S^{l}S_{l}}}%
,\qquad w^{i}j_{i}=0.  \label{a5.8}
\end{equation}

Let us take into account that the last term in the first equation (5.8) is
small with respect to the first term of Eq.(4.16). Then neglecting small
terms and taking into account Eq.(5.7), one obtains for the action
(4.15)--(4.18) in the quasi-uniform state
\begin{equation*}
\mathcal{A}_{Dqu}[j,\varphi ,\kappa ,\mathbf{\xi }]=\int \{-m\rho \cos
\kappa -eA_{i}j^{i}-\hbar j^{i}[\partial _{i}\varphi +g_{\alpha }(\mathbf{%
\xi })\partial _{i}\xi ^{\alpha }]
\end{equation*}%
\begin{equation}
+\frac{\hbar j^{i}}{{2\rho (}\rho +j^{j}f_{j}{)}}\varepsilon
_{klsm}j^{k}\partial _{i}j^{l}f^{s}\xi ^{m}\}d^{4}x  \label{a5.9}
\end{equation}%
where $g_{\alpha }(\mathbf{\xi })$ is determined by Eq.(4.5).

Total derivatives $\rho ^{-1}j^{i}\partial _{i}$ of variables $j^{i}$, $%
\kappa $, $\mathbf{\xi }$ in the dynamic system $\mathcal{S}_{Dqu}$
described by the action (5.9) are determined mainly by dynamic equations.
Derivatives in the orthogonal directions are determined by the initial
conditions which must be such ones, that these derivatives were small enough.

Although the action (5.9) contains a non-classical variable $\kappa $, in
fact this variable is a constant. Indeed, a variation with respect to $%
\kappa $ leads to the dynamic equation
\begin{equation}
\frac{\delta \mathcal{A}_{Dqu}}{\delta \kappa }=m\rho \sin \kappa =0
\label{a5.10}
\end{equation}%
which has solutions
\begin{equation}
\kappa =n\pi ,\qquad n=\func{integer}  \label{a5.11}
\end{equation}%
Thus, the effective mass $m_{eff}=m\cos \kappa $ has two values
\begin{equation}
m_{eff}=m\cos \kappa =\pm m  \label{a5.12}
\end{equation}%
The value $m_{eff}=m>0$, $(\kappa ={\frac{1}{2}}n\pi )$ corresponds to a
minimum of the action (5.9), whereas the value $m_{eff}=-m<0$ corresponds to
a maximum. Apparently, $m_{eff}>0$ corresponds to a stable ensemble state,
and $m_{eff}<0$ does to unstable state.

Eliminating $\kappa $ by means of the substitution $k={\frac{1}{2}}n\pi $ in
Eq.(5.9), one obtains the action
\begin{equation*}
\mathcal{A}_{Dqu}[j,\varphi ,\mathbf{\xi }]=\int \{-m\rho -eA_{i}j^{i}-\hbar
j^{i}[\partial _{i}\varphi +g_{\alpha }(\mathbf{\xi })\partial _{i}\xi
^{\alpha }]
\end{equation*}%
\begin{equation}
+\frac{\hbar j^{i}}{{2\rho (}\rho +j^{j}f_{j}{)}}\varepsilon
_{klsm}j^{k}\partial _{i}j^{l}f^{s}\xi ^{m}\}d^{4}x  \label{a5.13}
\end{equation}

Let us introduce Lagrangian coordinates $\tau =\{\tau _{i}\}$, $i=0,1,2,3$
by means of relations
\begin{equation}
j^{i}=\frac{\partial D}{\partial \tau _{0,i}}\equiv \frac{\partial
(x^{i},\tau _{1},\tau _{2},\tau _{3})}{\partial (x^{0},x^{1},x^{2},x^{3})}%
,\qquad D\equiv \frac{\partial (\tau _{0},\tau _{1},\tau _{2},\tau _{3})}{%
\partial (x^{0},x^{1},x^{2},x^{3})};  \label{a5.14}
\end{equation}%
\begin{equation*}
\tau _{k,i}\equiv \partial _{i}\tau _{k},\qquad i,k=0,1,2,3
\end{equation*}%
Taking into account that
\begin{equation}
D^{-1}j^{i}\partial _{i}u=D^{-1}\frac{\partial D}{\partial \tau _{0,i}}%
\partial _{i}u=\frac{\partial (u,\tau _{1},\tau _{2},\tau _{3})}{\partial
(\tau _{0},\tau _{1},\tau _{2},\tau _{3})}=\frac{du}{d\tau _{0}}
\label{a5.15}
\end{equation}%
\begin{equation}
d^{4}x=D^{-1}d^{4}\tau =D^{-1}d\tau _{0}d\mathbf{\tau }  \label{a5.16}
\end{equation}%
\begin{equation}
j^{i}\partial _{i}\varphi =\frac{\partial (\varphi ,\tau _{1},\tau _{2},\tau
_{3})}{\partial (x^{0},x^{1},x^{2},x^{3})}  \label{at.17}
\end{equation}%
the action (5.13) can be rewritten in the Lagrangian coordinates in the form
\begin{equation}
\mathcal{A}_{Dqu}[x,\mathbf{\xi }]=\int \{-m\sqrt{\dot{x}^{i}\dot{x}_{i}}%
-eA_{i}\dot{x}^{i}+\hbar {\frac{(\dot{\mathbf{\xi }}\times \mathbf{\xi })%
\mathbf{z}}{2(1+\mathbf{\xi })}}+\hbar \frac{(\dot{\mathbf{x}}\times \ddot{%
\mathbf{x}})\mathbf{\xi }}{2\sqrt{\dot{x}^{s}\dot{x}_{s}}(\sqrt{\dot{x}^{s}%
\dot{x}_{s}}+\dot{x}^{0})}\}d^{4}\tau  \label{a5.18}
\end{equation}%
where the dot means the total derivative $d/d\tau _{0}$. $x=\{x^{i}\}$, $%
i=0,1,2,3$, $\mathbf{\xi }=\{\xi ^{\alpha }\}$, $\alpha =1,2,3$ are
considered as functions of the Lagrangian coordinates $\tau _{0}$, $\mathbf{%
\tau }=\{\tau _{1},\tau _{2},\tau _{3}\}$. $\mathbf{z}$ is a constant unit
3-vector. $A_{i}=A_{i}(x)$, $i=0,1,2,3$ are function of $x$. The term $%
j^{i}\partial _{i}\varphi $ is omitted, because it reduces to a Jacobian
(5.17) and does not contribute into dynamic equations.

The action (5.18) describes a statistical ensemble of deterministic dynamic
systems $\mathcal{S}_{dc}$. A state of each system $\mathcal{S}_{dc}$ is
described by the variables $x^{i}$, $\dot{x}^{i}$, $\ddot{x}^{i}$, $\mathbf{%
\xi }$. The variables $\mathbf{\xi }$ are connected with the spin by the
relation (3.23) which takes the form

\begin{equation}
\mathbf{\xi }=\mathbf{s}-\frac{(\mathbf{s\dot{x}})}{\dot{x}^{0}(\sqrt{\dot{x}%
^{i}\dot{x}_{i}}+\dot{x}^{0})}\mathbf{\dot{x}},\qquad \mathbf{s}=\mathbf{\xi
}+\frac{(\mathbf{\xi \dot{x}})}{\sqrt{\dot{x}^{i}\dot{x}_{i}}(\dot{x}^{0}+%
\sqrt{\dot{x}^{i}\dot{x}_{i}})}\mathbf{\dot{x}}  \label{a5.19}
\end{equation}%
Here $\mathbf{s}=\mathbf{S}/\rho $ is the unit spin 3-vector.

All this means that in the quasi-uniform approximation the dynamic system $%
\mathcal{S}_{D}$ is a statistical ensemble of some deterministic classical
systems $\mathcal{S}_{dc}$. The system $\mathcal{S}_{dc}$ should be treated
as a classical analog of the Dirac electron moving in a given
electromagnetic field. Under some conditions the classical Dirac electron
turns into a classical relativistic pointlike particle, but, in general, $%
\mathcal{S}_{dc}$ is a more complicated construction than a pointlike
particle. As far as under some conditions $\mathcal{S}_{D}$ is a statistical
ensemble of classical relativistic particles, one concludes that, generally,
the Dirac equation describes a statistical ensemble of charged quantum
particles (not a single particle). Indeed, at first, one knew only that the
system $\mathcal{S}_{D}$ relates to an electron in some way, but one did not
know whether $\mathcal{S}_{D}$ describes a single electron or a statistical
ensemble of electrons. One discovers that under some (quasi-uniform) initial
conditions the $\mathcal{S}_{D}$ is a statistical ensemble (of classical
systems $\mathcal{S}_{dc}$). It means that $\mathcal{S}_{D}$ is a
statistical ensemble in all other cases. But in the general case $\mathcal{S}%
_{D}$ cannot be an statistical ensemble of deterministic classical systems.
It means that $\mathcal{S}_{D}$ is a statistical ensemble of stochastic
systems.

As far as the quantum principles are not used, then one uses the statistical
principle formulated in the sec.1. This principle permits to determine mean
values of energy, momentum, angular momentum and other additive quantities
for a single electron. Indeed, dividing an additive quantity for $\mathcal{S}%
_{D}$ by the number of systems in the ensemble $\mathcal{S}_{D}$, one
obtains corresponding mean value for a single stochastic system.

Let us compare the quasi-uniform approximation with the quasi-classical
approximation which is obtained by tending $\hbar \rightarrow 0.$ Let us go
to the limit $\hbar \rightarrow 0$ in the action (4.15)-(4.18). One obtains
\begin{equation}
\mathcal{A}_{Dcl}[j,\Phi ]=\int [-m\sqrt{j^{l}j_{l}}-eA_{l}j^{l}-j^{i}%
\partial _{i}\Phi ]d^{4}x  \label{a5.20}
\end{equation}%
where%
\begin{equation}
\Phi =\hbar \varphi  \label{a5.21}
\end{equation}
The term $-\hbar j^{i}\partial _{i}\varphi $ of Eq. (4.16) can be saved at $%
\hbar \rightarrow 0$ by means of the substitution $\varphi \rightarrow \Phi
=\hbar \varphi $. But the last term of $\mathcal{L}_{cl}$ cannot be
conserved by a like substitution, because of the restriction (4.20). For
instance, due to Eq.(4.20) the substitution $\mathbf{\xi }\rightarrow
\mathbf{w}=\hbar ^{1/2}\mathbf{\xi }$ leads to $\mathbf{w}=0$ at $\hbar
\rightarrow 0$.

The action (5.20) describes only a part of extremals (solutions) of the
action (5.13), namely that part of them which does not contain spin
variables (3.23) and describes a potential solution, where the momentum
(2.12) forms a potential vector field.
\begin{equation}
p_{l}=\partial _{l}\Phi  \label{a5.22}
\end{equation}%
The quasi-uniform approximation (5.13) obtained by a proper choice of rather
smooth initial conditions is more realistic, than the quasi-classical
approximation (5.20) obtained in the limit $\hbar \rightarrow 0,$ because in
reality nobody can change the quantum constant $\hbar $. Thus, the
quasi-uniform approximation is more preferable as a classical approximation
of the dynamic system $\mathcal{S}_{D}$.

Thus, a usage of the quasi-uniform approximation permits to separate out a
classical part of the action (4.15)-(4.18). The quantum part $\mathcal{L}%
_{q}=\mathcal{L}_{q1}+\mathcal{L}_{q2}$ of the action (4.15) is determined
by Eqs.(4.17), (4.18). $\mathcal{L}_{q1}$ contains a specific variable $%
\kappa $ which can be treated neither as a current, nor as a Lagrangian
coordinate $\mathbf{\xi }$. Suppressing $\kappa $ (i.e. setting $\kappa
\equiv 0)$, the term $\mathcal{L}_{q1}$ vanishes. The term $\mathcal{L}_{q2}$
is a quantum term, in general, although it contains a classical part which
can be separated out by means of introduction of specific quantum variables.

Let us introduce new variables
\begin{equation}
\kappa ^{i}=q^{i}=\frac{j^{i}+\rho f^{i}}{\sqrt{2\rho (\rho +j^{s}f_{s})}}=%
\frac{j^{i}+\rho f^{i}}{\sqrt{(j^{s}+\rho f^{s})(j_{s}+\rho f_{s})}},\qquad
i=0,1,2,3;  \label{a5.23}
\end{equation}%
\begin{equation*}
\rho \equiv \sqrt{j^{l}j_{l}},
\end{equation*}%
using the Lagrangian multipliers $\lambda _{i}$. Then $\mathcal{L}_{q2}$ is
substituted by
\begin{equation*}
\mathcal{L}_{q2}^{\prime }=\hbar \rho \varepsilon _{iklm}(\partial _{\perp
}^{k}\kappa ^{l})\kappa ^{i}\nu ^{m}+{\frac{\hbar }{\rho }}j^{s}\varepsilon
_{iklm}q^{i}j^{k}\partial _{s}q^{l}\nu ^{m}+
\end{equation*}%
\begin{equation}
+\lambda _{i}[j^{i}+\rho f^{i}-\kappa ^{i}\sqrt{2\rho (\rho +j^{s}f_{s})}]
\label{a5.24}
\end{equation}%
\begin{equation}
\rho \equiv \sqrt{j^{l}j_{l}},\qquad \nu ^{m}=[\xi ^{m}-f^{m}(\xi
^{s}f_{s})],\qquad m=0,1,2,3  \label{a5.25}
\end{equation}%
where $\partial _{\perp }^{k}$ is defined by Eq.(5.4). Then variation of the
action with respect to $\kappa ^{i}$ leads to the dynamic equations
\begin{equation}
-\lambda _{i}\sqrt{2\rho (\rho +j^{s}f_{s})}+{\frac{\delta \mathcal{A}%
_{q2}^{\prime \prime }}{\delta \kappa ^{i}}}=0,  \label{a5.26}
\end{equation}%
\begin{equation}
\mathcal{A}_{q2}^{\prime \prime }[j,\xi ,\kappa ^{i}]=\int \hbar \rho
\varepsilon _{iklm}(\partial _{\perp }^{k}\kappa ^{i})\kappa ^{l}\nu
^{m}d^{4}x  \label{a5.27}
\end{equation}

Resolving the equation (5.26) with respect to $\lambda _{i}$ and
substituting the $\lambda _{i}$ into Eq.(5.24), one obtains instead of $%
\mathcal{L}_{q2}^{\prime }$
\begin{equation}
\mathcal{L}_{q3}=\hbar \rho \varepsilon _{iklm}(\kappa ^{l}\partial _{\perp
}^{k}\kappa ^{i}-q^{l}\partial _{\perp }^{k}\kappa ^{i}+\kappa ^{i}\partial
_{\perp }^{k}q^{l})\nu ^{m}+\frac{\hbar j^{i}}{{2\rho (}\rho +j^{j}f_{j}{)}}%
\varepsilon _{klsm}j^{k}\partial _{i}j^{l}f^{s}\xi ^{m}  \label{a5.28}
\end{equation}%
Now the action has the form
\begin{equation}
\mathcal{A}_{D}[j,\varphi ,\kappa ,\xi ,\kappa ^{i}]=\int (\mathcal{L}_{cl}+%
\mathcal{L}_{q1}+\mathcal{L}_{q3})d^{4}x  \label{a5.29}
\end{equation}%
Dynamic equations generated by actions (3.1), (4.15) and (5.29) are
equivalent. The quantities $\xi ^{\alpha }$ cannot be treated as Lagrangian
coordinates (numbers labelling systems of the ensemble), because they are
not constant along world lines of particles, and relations (2.10) do not
take place. According to Eq.(3.23) the quantities $\xi ^{\alpha }$ should be
treated as some functions of the spin $\mathbf{S}%
=\{2S^{23,0},2S^{31,0},2S^{12,0}\}=\{S^{1},S^{2},S^{3}\}$.

The variables $\kappa $, $\kappa ^{i}$, $i=0,1,2,3$ are special quantum
variables which are responsible for quantum effects described by the Dirac
equation. Indeed, let us set $\kappa \equiv 0$, $\kappa ^{i}\equiv 0$, $%
i=0,1,2,3$ in the action (5.29). Then the action (5.29) turns to Eq.(5.13)
with $g_{\alpha }(\mathbf{\xi })$ defined by Eq.(4.5). The action (5.29)
generates the dynamic equation
\begin{equation}
{\frac{\delta \mathcal{A}_{D}}{\delta \kappa ^{i}}}=\hbar \varepsilon
_{iklm}\{\rho \nu ^{m}\partial _{\perp }^{k}(q^{l}-\kappa ^{l})+\partial
_{\perp }^{\ast k}[\rho \nu ^{m}(q^{l}-\kappa ^{l})]\}=0  \label{a5.30}
\end{equation}%
where the operator $\partial _{\perp }^{\ast k}$ is defined by the relation
\begin{equation}
\partial _{\perp }^{\ast k}u=\partial ^{k}u-\partial _{s}({\frac{j^{k}j^{s}}{%
\rho ^{2}}}u)  \label{a5.31}
\end{equation}%
Resolving Eq.(\ref{a5.30}) with respect to $\kappa ^{i}$ and substituting
the $\kappa ^{i}$ into Eq.(5.29), one returns to the action (4.15)-(4.18).
The fact that the solution (5.23) of Eq.(5.30) is not sole is of no
importance, because Eq.(5.28) reduces to Eq.(4.18) by virtue of Eq.(5.30).
Indeed, convoluting Eq.(5.30) with $\kappa ^{i}$ and using the obtained
relation for eliminating $\kappa ^{i}$ from Eq.(5.24), one obtains Eq.(4.18).

Let us note that $\kappa ^l$ are not rigorous dynamic variables, because the
dynamic equations (5.30) for $\kappa ^l$ contain derivatives only along
spacelike directions orthogonal to $j^i$. Rather the introduction of $\kappa
^i$ is an invariant (with respect to a change of variables) way of
separating out the classical part of the action.

The concentrated dynamic system $\mathcal{S}_{dc}$ has eleven degrees of
freedom. It is associated with the distributed dynamic system $\mathcal{S}%
_{D}$. It is described by the action
\begin{equation}
\mathcal{A}_{dc}[x,\mathbf{\xi }]=\int \{-m\sqrt{\dot{x}^{i}\dot{x}_{i}}%
-eA_{i}\dot{x}^{i}+\hbar {\frac{(\dot{\mathbf{\xi }}\times \mathbf{\xi })%
\mathbf{z}}{2(1+\mathbf{\xi z})}}+\hbar \frac{(\dot{\mathbf{x}}\times \ddot{%
\mathbf{x}})\mathbf{\xi }}{2\sqrt{\dot{x}^{s}\dot{x}_{s}}(\sqrt{\dot{x}^{s}%
\dot{x}_{s}}+\dot{x}^{0})}\}d\tau _{0}  \label{b5.32}
\end{equation}%
with non-relativistically invariant Lagrangian. It is a very surprising fact
which needs a special investigation.

\section{Relativistical Invariance}

At our consideration of the relativistic invariance of the Dirac equation
written in hydrodynamic terms we shall follow the approach of Anderson [6]
with the modification that the definition of the relativistic covariance is
provided by an explicit reference to the quantities with respect to which
the dynamic equations are relativistically covariant. Let us consider a
simple example which is relevant to the Dirac equation.

One considers a system of differential equations consisting of the Maxwell
equations for the electromagnetic tensor $F^{ik}$ in some inertial
coordinates $x$

\begin{equation}
\partial _{k}F^{ik}=4\pi J^{i},\qquad \varepsilon _{iklm}g^{ij}\partial
_{j}F^{kl}=0  \label{a6.1}
\end{equation}%
and equations
\begin{equation}
m\frac{d}{d\tau }[(l_{k}\dot{q}^{k})^{-1}\dot{q}^{i}-{\frac{1}{2}}%
g^{ik}l_{k}(l_{j}\dot{q}^{j})^{-2}\dot{q}^{s}g_{sl}\dot{q}%
^{l}]=eF^{il}g_{lk}(q)\dot{q}^{k};\qquad i=0,1,2,3  \label{a6.2}
\end{equation}%
\begin{equation*}
\dot{q}^{k}\equiv \frac{dq^{k}}{d\tau }
\end{equation*}%
where $q^{i}=q^{i}(\tau )$, $i=0,1,2,3$ describe coordinates of a pointlike
charged particle as functions of a parameter $\tau $, $l_{i}$ is a constant
timelike unit vector,
\begin{equation}
g^{ik}l_{i}l_{k}=1;  \label{a6.3}
\end{equation}%
and the speed of the light $c=1$.

This system of equations is relativistically covariant with respect to
quantities $q^i$, $F^{ik}$, $J^i$, $l_i$, $g_{ik}$, i.e. it does not change
its form at any infinitesimal Lorentz transformation

\begin{equation}
x^{i}\rightarrow \tilde{x}^{i}=x^{i}+\omega _{.k}^{i}x^{k}+o(\omega );\qquad
\omega _{.k}^{i}=g^{il}\omega _{lk};\qquad \omega _{ik}=-\omega _{ki}
\label{a6.4}
\end{equation}%
which is accompanied by corresponding transformation of quantities $q^{i}$, $%
F^{ik}$, $J^{i}$, $l_{i}$, $g_{ik}$,
\begin{equation}
q^{i}(\tau )\rightarrow \tilde{q}^{i}(\tau )=\frac{\partial \tilde{x}^{i}}{%
\partial x^{k}}q^{k}(\tau )=q^{i}+\omega _{.k}^{i}q^{k}+o(\omega )
\label{a6.5}
\end{equation}%
\begin{equation}
F^{ik}(x)\rightarrow \tilde{F}^{ik}(\tilde{x})=\frac{\partial \tilde{x}^{i}}{%
\partial x^{l}}\frac{\partial \tilde{x}^{k}}{\partial x^{j}}%
F^{lj}(x)=F^{ik}+\omega _{.l}^{i}F^{lk}+\omega _{.l}^{k}F^{il}+o(\omega )
\label{a6.6}
\end{equation}%
\begin{equation}
J^{i}(x)\rightarrow \tilde{J}^{i}(\tilde{x})=\frac{\partial \tilde{x}^{i}}{%
\partial x^{k}}J^{k}(x)=J^{i}+\omega _{.k}^{i}J^{k}+o(\omega )  \label{a6.7}
\end{equation}%
\begin{equation}
l_{i}\rightarrow \tilde{l}_{i}=\frac{\partial x^{k}}{\partial \tilde{x}^{i}}%
l_{k}=l_{i}+\omega _{i}^{.k}l_{k}+o(\omega ),\qquad \omega
_{i}^{.k}=g^{kl}\omega _{il}  \label{a6.8}
\end{equation}%
\begin{equation}
g_{ik}(x)\rightarrow \tilde{g}_{ik}(\tilde{x})=\frac{\partial x^{l}}{%
\partial \tilde{x}^{i}}\frac{\partial x^{j}}{\partial \tilde{x}^{k}}%
g_{lj}(x)=g_{ik}+\omega _{i}^{.l}g_{lk}+\omega _{k}^{.l}g_{il}+o(\omega )
\label{a6.9}
\end{equation}

The reference to the quantities $q^i$, $F^{ik}$, $J^i$, $l_i$, $g_{ik}$
means that all these quantities are considered as formal dependent
variables, when one compares the form of dynamic equations written in two
different coordinate systems. For instance, if a reference to $J^i$ is
omitted in the formulation of the relativistic covariance, it means that $%
J^i $ are considered as some functions of the coordinates $x$. If $J^i\neq 0$%
, then according to Eq.(6.7) $J^i$ and $\tilde{J}^i$ are different functions
of the arguments $x$ and $\tilde{x}$ respectively, and the first equation
(6.1) has different form in different coordinate systems. In other words,
the dynamic equations (6.1)--(6.2) are not relativistically covariant with
respect to quantities $q^i$, $F^{ik}$, $l_i$, $g_{ik}$. Thus, for the
relativistic covariance it is important both the laws of transformation
(6.5)--(6.9) and how each of quantities is considered as a formal variable,
or as some function of coordinates.

Following Anderson [6] we divide the quantities $q^i$, $F^{ik}$, $J^i$, $l_i$%
, $g_{ik}$ into two parts: dynamical objects (variables) $q^i$, $F^{ik}$ and
absolute objects $J^i$, $l_i$, $g_{ik}$. By definition of absolute objects
they have the same value for all solutions of the dynamic equations, whereas
dynamic variables are different, in general, for different solutions. If the
dynamic equations are written in the relativistically covariant form, their
symmetry group (and a compatibility with the relativity principles) is
determined by the symmetry group of the absolute objects $J^i$, $l_i$, $%
g_{ik}$.

Let for simplicity $J^i\equiv 0$. A symmetry group of the constant timelike
vector $l_i$ is a group of rotations in the $3$-plane orthogonal to the
vector $l_i$. The Lorentz group is a symmetry group of the metric tensor $%
g_{ik}=$diag $\{1,-1,-1,-1\}$. Thus, the symmetry group of all absolute
objects $l_i$, $g_{ik}$, $J^i\equiv 0$ is a subgroup of the Lorentz group
(rotations in the $3$-plane orthogonal to $l_i$). As far as the symmetry
group is a subgroup of the Lorentz group and does not coincide with it, the
system of equations (6.1)--(6.2) is non-relativistic (incompatible with the
relativity principles).

Of course, the compatibility with the relativity principles does not depend
on the fact with respect to which quantities the relativistic covariance is
considered. For instance, let us consider a covariance of Eqs. (6.1), (6.2)
with respect to quantities $q^{i}$, $F^{ik}$, $J^{i}\equiv 0$. It means that
now $l_{i}$ are to be considered as functions of $x$ (in the given case
these functions are constants), because a reference to $l_{i}$ as a formal
variables is absent. After the transformations (6.4)--(6.9) the equation
(6.2) takes the form
\begin{equation}
m\frac{d}{d\tau }[(\tilde{l}_{k}\frac{d\tilde{q}}{d\tau }^{k})^{-1}\frac{d%
\tilde{q}^{i}}{d\tau }-\frac{\tilde{l}^{i}}{2}(\tilde{l}_{k}\frac{d\tilde{q}%
^{k}}{d\tau })^{-2}\frac{d\tilde{q}}{d\tau }^{s}\frac{d\tilde{q}_{s}}{d\tau }%
]=e\tilde{F}_{.k}^{i}\frac{d\tilde{q}^{k}}{d\tau }  \label{a6.10}
\end{equation}

Here $\tilde{l} _i$ are considered as functions of $\tilde{x}$. But $\tilde{l%
}_i$ are other functions of $\tilde{x}$, than $l_i$ of $x$ (other numerical
constants $\tilde{l}_k=l_{j}\partial x^{j}/\partial\tilde{x}^{k}$ instead of
$l_{k}$), and equations (6.2) and (6.10) have different forms with respect
to quantities $q^i$, $F^{ik}$, $J^i\equiv 0$. It means that Eq.(6.2) is not
relativistically covariant with respect to $q^i$, $F^{ik}$, $J^i\equiv 0$,
although it is relativistically covariant with respect to $q^i$, $F^{ik}$, $%
l_i$, $J^i\equiv 0$.

Setting $l_{i}=\{1,0,0,0\}$, $t=q^{0}(\tau )$ in Eq.(6.2), one obtains
\begin{equation*}
m\frac{d^{2}q}{dt^{2}}^{\alpha }=eF_{.0}^{\alpha }+eF_{.\beta }^{\alpha }%
\frac{dq}{dt}^{\beta },\qquad i=\alpha =1,2,3;
\end{equation*}%
\begin{equation}
\frac{m}{2}\frac{d}{dt}(\frac{dq}{dt}^{\alpha }\frac{dq}{dt}^{\alpha
})=eF_{.0}^{\alpha }\frac{dq}{dt}^{\alpha },\qquad i=0.  \label{a6,11}
\end{equation}%
It is easy to see that this equation describes a non-relativistic motion of
a charged particle in a given electromagnetic field $F^{ik}$. The fact that
the equations (6.2) or (6.11) are non-relativistic is connected with the
space-time split into space and time that is characteristic for Newtonian
mechanics. This space-time split is described in different ways in Eqs.
(6.2) and (6.11). It is described by the constant timelike vector $l_{i}$ in
Eq.(6.2). In the equation (6.11) the space-time split is described by a
special choice of the coordinate system whose time axis is directed along
the vector $l^{i}$.

Thus, a relativistic covariance in itself does not mean a compatibility with
the special relativity principles. It is important with respect to which
quantities the dynamic equations are relativistically covariant

There is something like that in the case of the Dirac equation. The Dirac
equation (1.1) is relativistically covariant with respect to variables $\psi
, A_{i}$, which are transformed at the Lorentz transformation (6.4) as
follows

\begin{equation}
A_{i}(x)\rightarrow \tilde{A}_{i}(\tilde{x})=\frac{\partial x^{k}}{\partial
\tilde{x}^{i}}A_{k}(x)=A_{i}+\omega _{i}^{.k}A_{k}+o(\omega )  \label{a6.12}
\end{equation}%
\begin{equation}
\psi (x)\rightarrow \tilde{\psi}(\tilde{x})=\exp ({\frac{1}{4}}\gamma
^{ik}\omega _{ik})\psi (x)+o(\omega ),  \label{a6.13}
\end{equation}%
\begin{equation*}
\gamma ^{i}\rightarrow \tilde{\gamma ^{i}}=\gamma ^{i},\qquad i=0,1,2,3.
\end{equation*}%
A direct physical meaning of the variables $\psi $ is unclear. Only such
quantities as the current $j^{i}$ and spin (1.6) have a direct physical
meaning. The action (4.15)-(4.18) is a result of the Dirac action (3.1)
transformation to variables $j^{i},S^{i}$ having a physical meaning.

Let us replace the relation (3.23) by the relation

\begin{equation}
\xi ^{l}=\rho ^{-1}[S^{l}-\frac{j^{l}S^{k}f_{k}}{\rho +j^{s}f_{s}}],\qquad
l=0,1,2,3;  \label{a6.14}
\end{equation}%
where the vector $f^{l}$ is defined by Eq.(4.7). According to relations
(1.4) the equation (6.14) is equivalent to Eq.(3.23), if the first relation
(4.7) takes place.

Let $f^{k}$ be transformed as a vector, then according to Eq.(6.14) $\xi
^{l} $ is a pseudo-vector, because $j^{l}$ is a vector, and $S^{l}$ is a
pseudo-vector. Substituting Eq.(6.14) into Eqs.(5.29), (4.16)-(4.17),
(5.28), one obtains the action for the Dirac equation in terms of variables $%
j,S,\kappa ,\varphi ,\kappa ^i$

\begin{equation}
\mathcal{A}_{D}[j,S,\kappa ,\varphi ,\kappa ^{i}]=\int (\mathcal{L}%
_{cl}^{\prime }+\mathcal{L}_{q1}+\mathcal{L}_{q3})d^{4}x  \label{a6.15}
\end{equation}%
\begin{equation}
\mathcal{L}_{cl}^{\prime }=-m\rho -ej^{i}A_{i}-\hbar j^{i}(\partial
_{i}\varphi +\varepsilon _{mjkl}\mu ^{j}\partial _{i}\mu ^{k}f^{m}z^{l})
\label{a6.16}
\end{equation}%
where $\mathcal{L}_{q1}$ and $\mathcal{L}_{q3}$ are defined by Eqs.(4.17),
(5.28), $\mu ^{j}$ is defined by Eqs. (4.12), (4.11), (6.14), and $\nu ^{i}$
is defined by Eq.(4.11)
\begin{equation}
\mu ^{k}=[w^{k}-\frac{j^{k}w^{m}f_{m}}{\rho +j^{s}f_{s}}][2(1+w^{l}z_{l}-%
\frac{j_{l}z^{l}w^{m}f_{m}}{\rho +j^{s}f_{s}})]^{-1/2},\qquad w^{i}\equiv
\frac{S^{i}}{\sqrt{-S^{l}S_{i}}}  \label{a6.17}
\end{equation}%
\begin{equation}
\nu ^{k}=w^{k}-\frac{j^{k}w^{m}f_{m}}{\rho +j^{s}f_{s}}  \label{a6.18}
\end{equation}%
For obtaining dynamic equations the action (6.15) is varied with respect to
variables $j^{i}$, $S^{i}$, $\kappa ^{i}$, $\varphi $, $\kappa $ under
additional constraints (1.4). Vector $f^{i}$ and pseudo-vector $z^{i}$ are
constant and satisfy the conditions (4.8).

The obtained dynamic equations are equivalent to the Dirac equation (1.1).
They are relativistically covariant with respect to quantities $j^{i}$, $%
S^{i}$, $\kappa ^{i}$, $\varphi $, $\kappa $, $A_i$, $f^{i}$, $z^{i}$. But
they are not relativistically covariant, generally, with respect to
variables $j^{i}$, $S^{i}$, $\kappa ^{i}$, $\varphi $, $\kappa $ $A_i$. It
is easy to verify, setting values (4.7) of $f^{i}$, $z^{i}$ into the action
(6.15)-(6.18).

Thus, a relativistical covariance of dynamic equations described in terms of
$A_{i}$, $j^{i}$, $S^{i}$, $\kappa ^{i}$, $\varphi $, $\kappa $ depends
essentially on introduction of additional quantities $f^{i},z^{i}$. These
quantities are constant. They arise from the zero divisor (3.5) which can be
written in the covariant form%
\begin{equation}
\Pi =\frac{1}{2}\left( 1+f_{l}\gamma ^{l}\right) \frac{1}{2}\left( 1-i\gamma
_{5}f_{i}z_{k}\gamma ^{ik}\right)  \label{a6.19}
\end{equation}%
where $f^{l}$ is a constant timelike unit vector, and $z^{l}$ is a constant
unit pseudo-vector.

It is easy to verify that under the conditions (4.8) the factors of the
matrix (6.19) commute, and each factor is a zero divisor. Using designations
(3.2), one obtains that under conditions (4.7) the expression (6.19)
coincides with Eq.(3.5). Thus, variables $f^{l},z^{l}$ are present in a
hidden form inside the zero divisor $\Pi $ of the expression (3.4) for the
wave function. When one says that the Dirac equation (1.1) is
relativistically covariant with respect to variables $A_{i},\psi $, one
implies also the absolute objects $f^{l},z^{l}$, hidden inside $\psi $. It
is impossible to construct a proper zero divisor without using absolute
objects.

Indeed, if all components of the wave function $\psi $ are given,
the relations (1.2) determine the quantities $j^l$ and $S^l$
uniquely. The quantities $\varphi $, $\kappa $, are determined to
within an additive constant and to within an additive $(2\pi
)$fold constant respectively. The quantities do not depend on the
form of the zero divisor $\Pi $. If the quantities $j^l$, $S^l$,
$\varphi $, $\kappa $ are given, the wave function components are
determined by the relation (3.25) with the zero divisor determined
by Eq.(3.5). The wave function components depend on parameters
determining the form of the zero divisor. In other words, the
transformation
from the variables $j^l$, $S^l$, $\varphi $, $\kappa $ to the wave function $%
\psi $ is not single-valued. Substitution of $\psi $ from Eq.(3.4) into
Eq.(1.1) leads to appearance of constant parameters $f^k$, $z^k$ in dynamic
equations, provided they are written in the relativistically covariant form.

A common continuous symmetry group of both vectors $f^k$ and $z^k$ is a
group of rotation inside the 2-plane orthogonal to the bivector $%
f^kz^l-f^lz^k$. The Lorentz group is not a symmetry group of the dynamic
equations. If the peculiar directions determined by parameters $f^k$, $z^k$
are not fictitious, then the dynamic equations generated by the action
(6.15) are incompatible with the relativity principles.

The variable $f^{l}$ describes the space-time split into space and time,
what is characteristic for the Newtonian mechanics. In this sense the Dirac
equation is not more relativistic, than the non-relativistic equation (6.2).

\section{Concentrated Dynamic System Associated with the Dirac Equation}

Let us return to an investigation of the dynamic system $\mathcal{S}_{dc}$.
The action (5.32) can be presented in the relativistically covariant form as
follows
\begin{equation*}
\mathcal{A}_{dc}[x,\xi ]=\int \{-m\sqrt{\dot{x}^{i}\dot{x}_{i}}-eA_{i}\dot{x}%
^{i}-\hbar \frac{\varepsilon _{iklm}\xi ^{i}\dot{\xi}^{k}f^{l}z^{m}}{2(1-\xi
^{s}z_{s})}
\end{equation*}%
\begin{equation}
+\hbar \frac{\varepsilon _{iklm}\dot{x}^{i}\ddot{x}^{k}f^{l}\xi ^{m}}{2\sqrt{%
\dot{x}^{s}\dot{x}_{s}}(\dot{x}^{l}f_{l}+\sqrt{\dot{x}^{l}\dot{x}_{l}})}%
\}d\tau  \label{a7.1}
\end{equation}%
where $z^{i}$ and $f^{i}$ satisfy Eqs.(4.8), and the dot means a
differentiation with respect to invariant parameter $\tau $. $\xi ^{i}$ is a
unit spacelike vector orthogonal to the vector $f^{i}$. It can be presented
in the form
\begin{equation}
\xi ^{i}=\frac{\eta ^{i}-f^{i}\eta ^{s}f_{s}}{\sqrt{-\eta ^{l}\eta
_{l}+(\eta ^{l}f_{l})^{2}}}  \label{a7.2}
\end{equation}%
where $\eta ^{i}$ is an arbitrary spacelike vector.

It is easy to verify that
\begin{equation}
\delta \xi ^{i}=\frac{(\delta _{l}^{i}+\xi ^{i}\xi _{l})(\delta
_{k}^{l}-f^{l}f_{k})\delta \eta ^{k}}{\sqrt{-\eta ^{s}\eta _{s}+(\eta
^{s}f_{s})^{2}}}  \label{a7,3}
\end{equation}%
Then varying Eq.(7.1) with respect to $\xi ^{i}$ and using that $\delta \eta
^{i}$ is an arbitrary vector, one obtains dynamic equations
\begin{equation}
{\frac{\delta \mathcal{A}_{dc}}{\delta \xi ^{i}}}(\delta _{l}^{i}+\xi
^{i}\xi _{l})(\delta _{k}^{l}-f^{l}f_{k})={\frac{\delta \mathcal{A}_{dc}}{%
\delta \xi ^{i}}}(\delta _{k}^{i}+\xi ^{i}\xi _{k}-f^{i}f_{k})=0,\qquad
k=0,1,2,3  \label{a7.4}
\end{equation}%
There are only two independent dynamic equations among the equations (7.4),
because a convolution of Eq.(7.4) with vectors $\xi ^{k}$ and $f^{k}$
converts them into identities.

The action (7.1) is invariant with respect to a change of the invariant
parameter $\tau $
\begin{equation}
\tau \rightarrow \tilde{\tau}=f(\tau )  \label{a7.5}
\end{equation}%
where $f$ is an arbitrary monotone function. In particular, $\tau $ can be
chosen in such a way that
\begin{equation}
\dot{x}^{i}\dot{x}_{i}=1  \label{a7.6}
\end{equation}%
In this case $\tau $ can be interpreted as a proper time.

Let us written down the dynamic equations in the case of absence of the
electromagnetic field $A_{i}=0.$

\begin{equation*}
{\frac{\delta \mathcal{A}_{dc}}{\delta x^{i}}}=\frac{d}{d\tau }\{\frac{m\dot{%
x}_{i}}{\sqrt{\dot{x}^{s}\dot{x}_{s}}}-\hbar \varepsilon _{iklm}[Qf^{l}\ddot{%
x}^{k}\xi ^{m}+\frac{1}{2}\dot{x}^{k}\frac{d}{d\tau }(Qf^{l}\xi ^{m})]-
\end{equation*}%
\begin{equation}
-{\frac{\hbar }{2}}\varepsilon _{jklm}f^{l}\dot{x}^{j}\ddot{x}^{k}\xi ^{m}{%
\frac{\partial Q}{\partial \dot{x}^{i}}}\}=0,\qquad i=0,1,2,3  \label{a7.7}
\end{equation}%
\begin{equation*}
\delta \xi ^{i}:\qquad -\{\varepsilon _{sklm}[Pf^{l}\dot{\xi}^{k}z^{m}+\frac{%
1}{2}\xi ^{k}\frac{d}{d\tau }(Pf^{l}z^{m})+\frac{1}{2}Q\dot{x}^{k}\ddot{x}%
^{l}f^{m}]+
\end{equation*}%
\begin{equation}
\frac{1}{2}\varepsilon _{jklm}f^{l}\xi ^{j}\dot{\xi}^{k}z^{m}\frac{\partial P%
}{\partial \xi ^{s}}\}(\delta _{i}^{s}+\xi ^{s}\xi _{i}-f^{s}f_{i})=0
\label{a7.8}
\end{equation}%
where $P$ and $Q$ are defined by the relations

\begin{equation}
Q=Q(\dot{x})=[\sqrt{\dot{x}^{s}\dot{x}_{s}}(\dot{x}^{l}f_{l}+\sqrt{\dot{x}%
^{s}\dot{x}_{s}})]^{-1}  \label{a7.9}
\end{equation}%
\begin{equation}
P=P(\xi )=(1-z^{l}\xi _{l})^{-1}  \label{a7.10}
\end{equation}%
Equation (7.7) is integrated in the form
\begin{equation}
{\frac{m\dot{x}_{i}}{\sqrt{\dot{x}^{s}\dot{x}_{s}}}}-\hbar \varepsilon
_{iklm}[Qf^{l}\ddot{x}^{k}{\xi }^{m}+{\frac{1}{2}}\dot{x}^{k}{\frac{d}{d\tau
}}(Qf^{l}\xi ^{m})]-{\frac{\hbar }{2}}\varepsilon _{jklm}f^{l}\dot{x}^{j}%
\ddot{x}^{k}{\xi }^{m}{\frac{\partial Q}{\partial \dot{x}^{i}}}=p_{i}
\label{a7.11}
\end{equation}%
\begin{equation*}
p_{i}=\text{const},\qquad i=0,1,2,3;
\end{equation*}

Let $\tau $ be the proper time, and the condition (7.6) takes place.
Convoluting Eq.(7.11) with $f^{i}$ and eliminating the last term in lhs of
Eq.(7.11), one obtains
\begin{equation}
-\varepsilon _{iklm}[\ddot{x}^{k}f^{l}\xi ^{m}+{\frac{1}{2}}\dot{x}^{k}f^{l}%
\dot{\xi}^{m}-\frac{\dot{x}^{k}f^{l}\xi ^{m}(\ddot{x}^{j}f_{j})}{2(1+\dot{x}%
^{s}f_{s})}]=(\delta _{i}^{s}-f_{i}f^{s})(u_{s}-\dot{x}_{s})\frac{1+\dot{x}%
^{l}f_{l}}{\lambda },  \label{a7.12}
\end{equation}%
\begin{equation*}
\lambda =\frac{\hbar }{m},\qquad u_{i}=\frac{p_{i}}{m},\qquad i=0,1,2,3
\end{equation*}%
One can see from Eqs.(7.8), (7.12) that only directions orthogonal to the
timelike vector $f^{i}$ are essential.

Let us choose $f^{i}$ in the form (4.7). Then using designations
\begin{equation}
x^{i}=\{x^{0},\mathbf{x}\},\qquad \xi ^{i}=\{0,\mathbf{\xi }\},\qquad
z^{i}=\{0,\mathbf{z}\},\qquad u^{i}=\{u^{0},\mathbf{u}\},  \label{a7,13}
\end{equation}%
Eqs. (7.8), (7.12) reduce to the form
\begin{equation}
\lbrack -\mathbf{\dot{\mathbf{\xi }}}+{\frac{(\mathbf{z}\dot{\mathbf{\xi }})%
}{2(1+\mathbf{z\xi })}}\mathbf{\xi }]\times \mathbf{z}+{\frac{\mathbf{\xi }(%
\dot{\mathbf{\xi }}\times \mathbf{z})}{2(1+\mathbf{z\xi })}}\mathbf{z}+\frac{%
(1+\mathbf{z\xi })}{2(1+\dot{x}^{0})}\mathbf{\dot{x}\times \ddot{x}}=C%
\mathbf{\xi }  \label{a7.14}
\end{equation}%
\begin{equation}
{\frac{d}{d\tau }}{(}\frac{\mathbf{\dot{x}}}{\sqrt{1+\dot{x}_{0}}}{)}\times
\mathbf{\xi }+\frac{\mathbf{\dot{x}}\times \dot{\mathbf{\xi }}}{2\sqrt{1+%
\dot{x}_{0}}}=\frac{\sqrt{1+\dot{x}_{0}}}{\lambda }{(\mathbf{\dot{x}}-%
\mathbf{u})}  \label{a7.15}
\end{equation}%
where $C$ is some indefinite function of $\tau $.

According to Eqs.(7.2), (4.7)
\begin{equation}
\mathbf{\xi }^{2}=1,\qquad \mathbf{\xi }\dot{\mathbf{\xi }}=0.  \label{a7.16}
\end{equation}%
Using these relations, the equation (7.14) reduces to the form (see Appendix
B)
\begin{equation}
\dot{\mathbf{\xi }}=-\frac{(\mathbf{\dot{x}\times \ddot{x}})}{1+\dot{x}_{0}}%
\times \mathbf{\xi }  \label{a7.17}
\end{equation}%
which does not contain the vector \textbf{z}. It means that $\mathbf{z}$
determines a fictitious direction in the space-time. Note that $\mathbf{z}$
in the action (4.15) for the system $\mathcal{S}_{D}$ is fictitious also,
because the term containing $\mathbf{z}$ is the same in the actions (4.15)
and (5.13) for $\mathcal{S}_{D}$ and $\mathcal{S}_{Dqu}$ respectively.

Eq.(7.17) can be written in the relativistically covariant form
\begin{equation}
\dot{\xi}_{i}=-\frac{\varepsilon _{iklm}\xi ^{l}f^{m}\varepsilon
_{.k^{\prime }l^{\prime }m^{\prime }}^{k}\dot{x}^{k^{\prime }}\ddot{x}%
^{l^{\prime }}f^{m^{\prime }}}{1+f_{s}\dot{x}^{s}}  \label{a7.18}
\end{equation}

The system of Eqs.(7.12), (7.18) is relativistically covariant with respect
to the quantities $x^{i}$, $\xi ^{i}$, $p_{i}$, $f^{i}$ considered as
4-vectors $(x^{i}$, $p_{i}$, $f^{i})$ and pseudo-vectors $(\xi ^{i})$. But
it is not relativistically covariant with respect to the dynamic variables $%
x^{i}, \xi ^{i}, p_{i}$. Under the Lorentz transformation (6.4) the set $%
\mathcal{S}_f$ of all solutions $\{x^{i}, \xi ^{i}, p_{i}\}$ of Eqs.(7.12),
(7.18) at fixed parameters $f^{i}$ is transformed into another set $\mathcal{%
S}_{\tilde{f}}$ of solutions $\{\tilde{x}^{i}, \tilde{\xi }^{i}, \tilde{p}%
_{i}\}$. But the set $\mathcal{S}=\{\mathcal{S}_{f}\}$ of all sets $\mathcal{%
S}_{f}$ is transformed into itself, provided $f^{i}$ satisfies Eq.(4.8).

Formally the integration constants $p_{i}$ can be considered as some
parameters of the system of differential equations (7.18), (7.12). At the
same time the parameters $p_{i}$, $f^{i}$ of Eqs. (7.18), (7.12) can be
considered as some integration constants of some system $\mathcal{D}$ of
differential equations which is relativistically covariant with respect to
its dynamic variables. The set of all solutions of $\mathcal{D}$ can be
obtained as a result of the Lorentz transformation (6.4) of the set $%
\mathcal{S}_{p}$ of all solutions with fixed values of $p_{i}$, or as that
of the set $\mathcal{S}_{f}$ of all solutions with fixed values of $f^{i}$.

It is sufficient to investigate solutions $\mathcal{S}_{f_0}$ of equations
(7.12), (7.18) at $f^i_{(0)}=\{1,0,0,0\}$, because solutions at other values
of $f^i$ can be obtained from solutions at $f^i=f^i_{(0)}$ by means of a
proper Lorentz transformation.

Eq.(7.17) describes a rotation of the unit 3-vector $\mathbf{\xi }$ around
the vector $\mathbf{\dot{x}\times \ddot{x}}$ with the angular frequency
\begin{equation}
\mathbf{\Omega }_{\xi }=\frac{\mathbf{\dot{x}\times \ddot{x}}}{1+\dot{x}_{0}}
\label{a7.19}
\end{equation}%
Eliminating $\dot{\mathbf{\xi }}$ by means of a substitution of Eq.(7.17)
into Eq.(7.15), and introducing a new variable $\mathbf{y}$ instead of $%
\mathbf{\dot{x}}$

\begin{equation}
\mathbf{y}=\frac{\mathbf{\dot{x}}}{\sqrt{1+\dot{x}^{0}}},\qquad \mathbf{\dot{%
x}}=\mathbf{y}\sqrt{2+\mathbf{y}^{2}},\qquad \dot{x}^{0}\equiv \sqrt{1+%
\mathbf{\dot{x}}^{2}},  \label{a7.20}
\end{equation}%
Eq.(7.15) reduces to the form

\begin{equation}
\lambda \mathbf{\dot{y}\times }(\mathbf{\xi }+{\frac{(\mathbf{\xi })}{2\sqrt{%
2+\mathbf{y}^{2}}}}\mathbf{y})=(\mathbf{y}\sqrt{2+\mathbf{y}^{2}}-\mathbf{u})%
\sqrt{2+\mathbf{y}^{2}}  \label{a7.21}
\end{equation}%
Let us introduce a dimensionless variable $\mathbf{w}$ which is supposed to
be small

\begin{equation}
\mathbf{w}=\mathbf{y}-\mathbf{b},\qquad \mathbf{b}=\mathbf{u}(1+\sqrt{1+%
\mathbf{u}^{2}})^{-1/2},\qquad \mid \mathbf{w\mid \ll }1  \label{a7.22}
\end{equation}%
Then Eq.(7.21) takes the form

\begin{equation}
\lambda \mathbf{\dot{w}\times a}=[G\mathbf{w}+\mathbf{b}(\mathbf{wb})]+O(%
\mathbf{w}^{2}),  \label{a7.23}
\end{equation}%
\begin{equation}
G=2+\mathbf{b}^{2},\qquad \mathbf{a}=\mathbf{\xi }+{\frac{(\mathbf{\xi b})}{2%
\sqrt{2+\mathbf{b}^{2}}}}\mathbf{b}  \label{a7.24}
\end{equation}%
It follows from Eq.(7.23) that \textbf{a} is orthogonal to rhs of Eq.(7.23),
or

\begin{equation}
\mathbf{wd}=0,\qquad \mathbf{d\equiv }G\mathbf{a}+(\mathbf{ba})\mathbf{b}
\label{a7.25}
\end{equation}%
Characteristic frequency for $\mathbf{w}$ is of the order of $\lambda ^{-1}$%
, whereas that for $\mathbf{\xi }$ is of the order of $\lambda ^{-1}\mid
\mathbf{w\mid \ll \lambda }^{-1}$. It means that, integrating Eq.(7.23), one
can consider $\mathbf{\xi }$ as a constant vector.

If $\mathbf{\xi }\parallel \mathbf{b},\ \mathbf{w}$ is orthogonal to $%
\mathbf{b}$, and Eq. (7.23) reduces to the form

\begin{equation}
\lambda \mathbf{\dot{w}}={\frac{G}{\mathbf{a}^{2}}}\mathbf{a\times w}+O(%
\mathbf{w}^{2}),\qquad \mathbf{w}^{2}=\text{const}  \label{a7.26}
\end{equation}%
which describes a rotation of the vector $\mathbf{w}$ with the angular
velocity $G\mathbf{a}/\mathbf{a}^{2}$.

If $\mathbf{\xi }\times \mathbf{b\neq }0,$ let us choose orthonormal vectors
$\mathbf{e}_{1},\mathbf{e}_{2},\mathbf{e}_{3}$ in such a way that $\mathbf{e}%
_{1},\mathbf{e}_{2}$ lie on the plane $\mathcal{P}$ spanned by vectors $%
\mathbf{\xi }$ and $\mathbf{b}$. Let the vector $\mathbf{e}_{1}$ be chosen
on $\mathcal{P}$ in such a way that $\mathbf{e}_{1}\parallel \mathbf{d}$.
Then one has the following decomposition of vectors $\mathbf{w},\mathbf{a},%
\mathbf{b}$

\begin{equation}
\mathbf{w}=w_{2}\mathbf{e}_{2}+w_{3}\mathbf{e}_{3},\qquad \mathbf{a}=a_{1}%
\mathbf{e}_{1}+a_{2}\mathbf{e}_{2},\qquad \mathbf{b}=b_{1}\mathbf{e}%
_{1}+b_{2}\mathbf{e}_{2}  \label{a7.27}
\end{equation}%
According to Eqs.(7.23), (7.25), (7.27) the first equation (7.25) reduces to
the form

\begin{equation}
w_{2}[a_{1}b_{1}b_{2}+a_{2}(G+b_{2}^{2})]=0.  \label{a7.28}
\end{equation}%
Substituting Eqs. (7.27) into Eq.(7.26), one obtains
\begin{equation*}
-\lambda \dot{w}_{3}a_{2}=b_{1}b_{2}w_{2}+O(\mathbf{w}^{2}),
\end{equation*}%
\begin{equation}
\lambda \dot{w}_{3}a_{1}=(G+b_{2}^{2})w_{2}+O(\mathbf{w}^{2}),  \label{a7.29}
\end{equation}%
\begin{equation*}
-\lambda \dot{w}_{2}a_{1}=Gw_{3}+O(\mathbf{w}^{2}),
\end{equation*}%
Due to Eq.(7.28) two first equations (7.29) are equivalent. The system
(7.29) can be easily solved. The solution has the form.
\begin{equation}
\mathbf{w}=A[\mathbf{e}_{2}\cos \phi +(1+b_{2}^{2}/G)^{1/2}\mathbf{e}%
_{3}\cos \phi ]+O(A^{2})  \label{a7.30}
\end{equation}%
where
\begin{equation*}
\mathbf{e}_{1}=\mathbf{d}/\mid \mathbf{d\mid },\qquad \mathbf{e}_{2}=[1-(%
\mathbf{e}_{1}\mathbf{\xi })^{2}]^{-1/2}\mathbf{e}_{1}\times \mathbf{\xi }%
,\qquad \mathbf{e}_{3}=\mathbf{e}_{1}\times \mathbf{e}_{2},
\end{equation*}%
\begin{equation}
\phi =\Omega _{x}(\tau -\tau _{(0)})+\phi _{(0)}  \label{a7.31}
\end{equation}%
\begin{equation}
\Omega _{x}={\frac{G}{\lambda a_{1}}}[1+b_{2}^{2}/G]^{1/2},\qquad a_{1}=(%
\mathbf{ae}_{1}),\qquad b_{2}=(\mathbf{be}_{2})  \label{a7.32}
\end{equation}%
Using Eqs.(7.20), (7.22), one obtains

\begin{equation*}
\mathbf{\dot{x}}=u+A[(Ge_{2}+\frac{b_{2}}{G}b)\cos \phi
+(G+b_{2}^{2})^{1/2}e_{3}\sin \phi ]+O(A^{2})
\end{equation*}%
\begin{equation}
\dot{x}^{0}=\sqrt{1+\mathbf{\dot{x}}^{2}}  \label{a7.33}
\end{equation}%
\begin{equation*}
\mathbf{x}=x_{(0)}+u(\tau -\tau _{(0)})+A\Omega _{x}^{-1}[(Ge_{2}+\frac{b_{2}%
}{G}b)\sin \phi -(G+b_{2}^{2})^{1/2}e_{3}\cos \phi ]+O(A^{2})
\end{equation*}%
\begin{equation}
x^{0}=x_{(0)}^{0}+\int\limits_{\tau _{(0)}}^{\tau }\sqrt{1+\mathbf{\dot{x}}%
^{2}(\tau ^{\prime })}d\tau ^{\prime }  \label{a7.34}
\end{equation}%
$A,\phi _{(0)},x_{(0)}^{0},\mathbf{x}_{(0)},\mathbf{u},\tau _{(0)}$ are
independent integration constants. $\mathbf{a},\mathbf{b},G$ and $\Omega
_{x} $ are determined by relations (7.22), (7.24), (7.32) through these
constants.

According to Eq.(7.19) the angular velocity $\Omega _{\xi }$ of the vector $%
\mathbf{\xi }$ rotation reduces to the form

\begin{equation}
\mathbf{\Omega }_{\xi }=\frac{A\Omega _{x}}{1+\dot{x}^{0}}[-u\times (Ge_{2}+%
\frac{b_{2}}{G}b)\sin \phi -(1+b_{2}^{2}G)^{1/2}\mathbf{u\times e}_{3}\cos
\phi ]+O(A^{2})  \label{a7.35}
\end{equation}%
$\mathbf{\Omega }_{\xi }$ oscillates rapidly around the mean value $<\mathbf{%
\Omega }_{\xi }>\simeq O(a^{2})$. It means that
\begin{equation}
\mathbf{\xi }=\mathbf{\xi }_{(0)}+O(A^{2})  \label{a7.36}
\end{equation}%
\begin{equation}
\mathbf{\xi }_{(0)}=\text{const},\qquad \mathbf{\xi }_{(0))}\mathbf{\xi }%
_{(0)}=1  \label{a7.37}
\end{equation}

In order to show that $f^{i}$ is not a fictitious parameter, let us consider
the exact solution of equations (7.12), (7.18) at $f^{i}$ defined by
Eq.(4.7) .
\begin{equation*}
x^{i}=\{\sqrt{1+a^{2}}\tau ,\ -a\Omega ^{-1}\sin \phi ,\ a\Omega ^{-1}\cos
\phi ,\ 0\};\qquad \xi ^{i}=\{0,0,0,1\}
\end{equation*}%
\begin{equation}
u^{i}=\{1,0,0,0\},\qquad \Omega ={\frac{1+\sqrt{1+a^{2}}}{\lambda }},\qquad
\phi =\Omega \tau +\phi _{(0)},\qquad a=\text{const}  \label{a7.38}
\end{equation}%
If $f^{i}$ is a fictitious, the system of equations (7.12), (7.18) is
relativistically covariant with respect to the vectors $x^{i},\xi ^{i},u^{i}$%
, and the vectors $\tilde{x}^{i}$, $\tilde{\xi}^{i}$, $\tilde{u}^{i}$,
obtained from Eqs.(7.38) by means of a Lorentz transformation are to form a
solution of Eqs.(7.12), (7.18).

In the coordinate system moving with the velocity $V=\tanh \vartheta $ in
the direction of the axis $x^{3}$ the vectors (7.38) have the form
\begin{equation*}
\tilde{x}^{i}=\{\tau \sqrt{1+a^{2}}\cosh \vartheta ,\ -a\Omega ^{-1}\sin
\phi ,\ a\Omega ^{-1}\cos \phi ,\ \tau \sqrt{1+a^{2}}\sinh \vartheta \};
\end{equation*}%
\begin{equation}
\tilde{\xi}^{i}=\{\sinh \vartheta ,\ 0,\ 0,\ \cosh \vartheta \};\qquad
\tilde{u}^{i}=\{\cosh \vartheta ,\ 0,\ 0,\ \sinh \vartheta \};  \label{a7.39}
\end{equation}%
\begin{equation*}
\Omega =\frac{1+\sqrt{1+a^{2}}}{\lambda }\qquad \phi =\Omega \tau +\phi
_{(0)},\qquad a=\text{const}
\end{equation*}%
It is easy to verify that (7.39) is not a solution of Eqs.(7.12), (7.18), if
$\vartheta \neq 0$ and $f^{i}$ is determined by the equation (4.7). (Of
course, Eq.(7.39) is a solution of Eqs. (7.12), (7.18), if $f^{i}=\tilde{f}%
^{i}=\{\cosh \vartheta ,0,0,\sinh \vartheta \}$). It means that the vector $%
f^{i}$ is an essential parameter, and the system of equations (7.12), (7.18)
is incompatible with the special relativity principle.

The fact that $z^{i}$ is fictitious, but $f^{i}$ is not, is explained,
apparently, by incomplete symmetry of the Dirac equation with respect to the
time and the space. The matrix $\gamma ^{0}$ is used for constructing the
Dirac conjugate spinor $\bar{\psi }=\psi ^{*}\gamma ^{0}$. It separates out
the matrix $\gamma ^0$ among matrices $\gamma ^i$

A space-time split generated by $\gamma ^{0}$ appears in the Space-Time
Algebra (STA) suggested by Hestenes [7]. STA is a kind of Clifford Algebra
describing space-time properties [8]. This space-time split is connected
with a use of the matrix $\gamma ^0$.

In the case, when the electromagnetic field does not vanish, the quantities $%
p_{i}$ determined by Eq.(7.11) are not constant. They satisfy the equations
\begin{equation}
\dot{p}_{i}\equiv m\dot{u}_{i}=eF_{ik}(x)\dot{x}^{k},\qquad F_{ik}\equiv
\partial _{i}A_{k}-\partial _{k}A_{i}.  \label{a7.40}
\end{equation}%
There are only three independent equations among them
\begin{equation}
m\mathbf{\dot{u}}=e\mathbf{E\dot{x}}^{0}+e[\mathbf{\dot{x}\times H}],\qquad
\mathbf{E}=\mathbf{E}(x^{0},\mathbf{x}),\qquad \mathbf{H}=\mathbf{H}(x^{0},%
\mathbf{x})  \label{a7.41}
\end{equation}%
where $\mathbf{E}=\{E_{\alpha }\}=\{-F_{\alpha 0}\}$, $\alpha =1,2,3$; $%
\mathbf{H}=\{H_{\alpha }\}=\{-{\frac{1}{2}}\varepsilon _{\alpha \beta \gamma
}F_{\beta \gamma }\}$, $\alpha =1,2,3.$ The system of equations (7.6),
(7.16), (7.17), (7.20), (7.41) is a complete system of dynamic equations for
the dynamic variables $\mathbf{\xi }$ ,$\mathbf{x}$, $x^{0}$, $\mathbf{u}$.

For not too strong electromagnetic field, when the Larmor frequency is much
less, than $\Omega _{x}$ ($eH/m\ll \Omega _{x}\simeq m/\hbar $, i.e. if $%
H\ll 10^{12}G),\mathbf{u}$ in Eq.(7.21) can be considered approximately as a
constant. Then the world line $x^{i}=x^{i}(\tau )$, $i=0,1,2,3$ describing a
solution of the system (7.6), (7.16), (7.17), (7.20), (7.21), (7.41) spirals
round the world line $X^{i}=X^{i}(\tau )$ which describes a motion of the
guiding center and satisfies the equation
\begin{equation}
m\ddot{X}^{i}=eF_{.k}^{i}(X)\dot{X}^{k},\qquad i=0,1,2,3  \label{a7.42}
\end{equation}

In general, a charged particle moving in such a way must intensively radiate
electromagnetic waves. As a result its world line is to approach to the
world line of the guiding center.

Such a helical motion of the electron associates with the classical model
[9] of the Schr\"odinger zitterbewegung [10]. Another approaches to the
interpretation of the zitterbewegung can be found in papers [11-14] and
references therein.

\section{Discussion}

It seems that $p_{i}$ and $\dot{x}^{i}$ should be interpreted respectively
as a 4-momentum and 4-velocity of the system $\mathcal{S}_{dc}$. But then
the system $\mathcal{S}_{dc}$ can be hardly interpreted as a pointlike
classical particle. There are two reasons. First, $p_{i}$ and $\dot{x}^{i}$
of a pointlike particle are connected by an algebraic relation of the kind
of $p_{i}=m\dot{x}_{i}$. For the dynamic system $\mathcal{S}_{dc}$ a
distinction between $y^{i}=\dot{x}^{i}{(1+\dot{x}^{0})}^{-1/2}$ and $%
u^{i}=p^{i}/m$ is described by the dynamic variable $w^{i}$ which is
introduced by Eq.(7.22). According to Eq.(7.21) this distinction is
proportional to the quantum constant $\hbar $. Second, it is very difficult
to understand, why the world line of a free particle is a helix spiralling
round the straight line
\begin{equation}
X^{i}=\frac{p^{i}}{m}(\tau -\tau _{(0)})+X_{(0)}^{i}.  \label{a8.1}
\end{equation}%
even in an absence of the electromagnetic field.

All this can be explained by means of the supposition that $x^i$ describes
an observable part of a complicated bound system whose center of inertia $%
X^i $ moves according to Eq.(8.1) in the case of $A_i\equiv 0$ and according
to Eq.(7.42) in the case of not too strong electromagnetic field. But under
such an interpretation the system $\mathcal{S}_{dc}$ is not a pointlike
particle with some inner degrees of freedom, but a dynamic system consisting
of a few particles interacting through a distance. But such an interaction
through a distance is incompatible with the relativity principle.

One can try to save the compatibility with the relativity by a consideration
of some distributed classical system $\mathcal{S}_{\phi }$ instead of the
concentrated dynamic system $\mathcal{S}_{dc}$. The system $\mathcal{S}_{dc}$
arises as an element of the statistical ensemble $\mathcal{S}_{Dqu}$. But an
one-parameter statistical ensemble of dynamic systems $\mathcal{S}_{dc}$ can
be also considered as an element of the statistical ensemble $\mathcal{S}%
_{Dqu}$. For instance, let us consider a set $\mathcal{S}_{\phi }$ of
systems $\mathcal{S}_{dc}$, having all similar integration constants $%
\{a,x_{(0)}^{0},\mathbf{x}_{(0)},\mathbf{u},\tau _{(0)},\mathbf{\xi }%
_{(0)}\} $ except of $\phi _{(0)}$. Let values of $\phi _{(0)}$ be
distributed uniformly among the systems $\mathcal{S}_{dc}$. (For instance,
in Eq.(7.38) $x^{i}$ is considered as a function of two variables $\tau $
and $\phi _{(0)}$ with fixed parameter $a$). Then $\mathcal{S}_{\phi }$ can
be considered as a ring. World lines of the ring particles form a
two-dimensional world tube in the space-time. Such a construction is more
symmetric, than helically moving particle. Besides such a ring does not
radiate electromagnetic waves. The statistical ensemble $\mathcal{S}_{Dqu}$
can be considered as consisted of rings $\mathcal{S}_{\phi }$ which are
classical distributed dynamic systems. Unfortunately, rings $\mathcal{S}%
_{\phi }$ and their two-dimensional world tubes are incompatible with the
concept of a pointlike particle.

One can try to associate the world tubes with the space-time properties (but
not with the structure of the dynamic system $\mathcal{S}_{\phi })$. In this
case it is necessary to admit that the distortion of the space-time does not
vanish, i.e. the characteristic geometric structures of the space-time are
three-dimensional world-tubes (but not straight lines) [2,15]. Then one can
hope to remove incompatibility between the concept of a pointlike particle
and the cylindrical space-time structures generated by the Dirac equation.
Note that in the distorted space-time the momentum of a particle and its
velocity are independent quantities which are connected by no algebraic
relation.

Thus, the well known dynamic system $\mathcal{S}_D$ described by the Dirac
equation has been investigated simply as a dynamic system without using any
additional suppositions. The obtained results can be summed as follows.

1. Describing the dynamic system $\mathcal{S}_D$ in terms of hydrodynamic
variables, one discovers that in the quasi-uniform approximation the $%
\mathcal{S}_D$ turns to a statistical ensemble $\mathcal{S}_{Dqu}$ of
classical systems $\mathcal{S}_{dc}$. It means that, in general, the dynamic
system $\mathcal{S}_D$ is a statistical ensemble of stochastic systems, and
it permits to separate out a classical part $\mathcal{L}_{Dqu}$ of the total
Lagrangian $\mathcal{L}_D$.

2. It is possible to separate out the classical part $\mathcal{L}_{Dqu}$ of
the Lagrangian by means of special quantum variables $\kappa $, $\kappa ^i$
responsible for quantum effects. Suppressing the quantum variables (setting $%
\kappa , \kappa ^i \equiv 0$), one transforms the Lagrangian $\mathcal{L}_D$
of the system $\mathcal{S}_D$ into its classical part $\mathcal{L}_{Dqu}$.

3. The transformation of the Dirac equation to hydrodynamic variables
separates out a peculiar direction in the space-time. This direction is
described by the timelike unit vector $f^k$. Existence of a preferred
direction in the space-time is incompatible with the special relativity
principle. It means that the Dirac equation is incompatible with the special
relativity principle and needs a modification for such a compatibility.

4. The dynamic system $\mathcal{S}_{dc}$ is a classical analog of the
quantum Dirac electron. $\mathcal{S}_{dc}$ cannot be interpreted as a
pointlike charged particle in the Minkowski space-time. Rather it is either
a complicated non-local dynamic system in the Minkowski space-time, or a
pointlike particle, but in a non-Riemannian (distorted) space-time.

\bigskip \centerline{\bf APPENDIX A}
\centerline{{\bf Classical Ensemble Described in
Lagrangian and Eulerian Coordinates} } \bigskip Let us show equivalency of
two form (2.1) and (2.2) of the action for the statistical ensemble of
classical systems. Let us introduce the time Lagrangian coordinate $\xi _{0}$%
, supposing that $t=t(\xi _{0})$, $x=\{t,\mathbf{x\}}=x(\xi )$, $\xi =\{\xi
_{0},\mathbf{\xi }\}$. Then the action (2.1) can be rewritten in the form
\begin{equation*}
A_{L}\left[ x\right] =\int \mathcal{L}\left( x,\dot{x}\right) d^{n+1}\mathbf{%
\xi ,\quad }\dot{x}\equiv \partial x/\partial \xi _{0},\quad d^{n+1}\mathbf{%
\xi =}d\xi _{0}d\mathbf{\xi }\eqno(A.1)
\end{equation*}

\begin{equation*}
\mathcal{L}(x,\dot{x})=L(x^{0},\mathbf{x},\mathbf{\dot{x}}/\dot{x}^{0})\dot{x%
}^{0}.\eqno(A.2)
\end{equation*}%
According to Eq.(A.2) the Lagrangian $\mathcal{L}(x,\dot{x})$ is a first
order homogeneous function of $\dot{x}$, i.e.
\begin{equation*}
\mathcal{L}(x,a\dot{x})=a\mathcal{L}(x,\dot{x}),\eqno(A.3)
\end{equation*}%
where $a$ is an arbitrary parameter.

Let the variational problem with the functional (A.1) take place. Let us
show that it is equivalent to the variational problem with the functional
(2.2).

Considering $\xi =\{\xi _i\}$, $i=0,1,\ldots n$ as functions of $x=\{x^i\}$,
$i=0,1,\dots n$, the variational problem with the action (A.1) can be
formulated as a variational problem with the action
\begin{equation*}
\mathcal{A}[\xi ]=\int \mathcal{L}(x,\frac{\partial J}{\partial \xi _{0,i}}%
)d^{n+1}x,\qquad \xi =\xi (x)\eqno (A.4)
\end{equation*}
where $J$ is a Jacobian
\begin{equation*}
J\equiv \frac{\partial (\xi _0,\xi _1,\ldots \xi _n)}{\partial
(x^0,x^1,\ldots x^n)}\equiv \det \parallel \xi _{i,k}\parallel ,
\end{equation*}
\begin{equation*}
\xi _{i,k}\equiv \partial \xi _i/\partial x^k,\qquad i,k=0,1,\ldots n\eqno %
(A.5)
\end{equation*}
and
\begin{equation*}
\frac{\partial J}{\partial \xi _{0,i}}\equiv \frac{\partial (x^i,\xi _1,\xi
_2,\ldots \xi _n)}{\partial (x^0,x^1,\ldots x^n)}\equiv {\frac{\partial x^i}{%
\partial \xi _0}}J,\qquad i=0,1,\ldots n\eqno (A.6)
\end{equation*}
Deriving Eq.(A.4) from Eq.(A.1), the relations (A.3), (A.6) were used.

In the variational problem with the action (A.4) the Lagrangian coordinates $%
\xi =\{\xi _{i}\}$, $i=0,1,\ldots n$ are dynamic variables considered as
functions of coordinates $x=\{x^{i}\}$, $i=0,1,\ldots n$.

Let us use designations
\begin{equation*}
j^i=\frac{\partial J}{\partial \xi _{0,i}},\qquad i=0,1,\ldots n\eqno (A.7)
\end{equation*}
Adding constraints (A.7) to the action (A.4), one does not change the
variational problem, because Eq.(A.4) does not contain variables $j^i$, and
Eq. (A.7) is not a constraint in reality. Let us introduce Eq.(A.7) into the
action (A.4) by means of Lagrange multpliers $p=\{p_i\}$, $i=0,1,\ldots n $.
$j^i$ and $p_i$, $i=0,1,\ldots n$ mean respectively a current density and a
momentum density in the space $\{t,\mathbf{x\}}$.

Then one obtains
\begin{equation*}
\mathcal{A}[j,p,\xi ]=\int [L(x,j)-p_i(j^i-\frac{\partial J}{\partial \xi
_{0,i}})]dx^{n+1}.\eqno (A.8)
\end{equation*}
Dynamic equations generated by the action (A.8) have the form

\begin{equation*}
{\frac{\delta \mathcal{A}}{\delta \xi _\alpha }}=-\partial _k(p_l\frac{%
\partial ^2J}{\partial \xi _{0,l}\partial \xi _{\alpha ,k}})=0,\qquad \alpha
=1,2,\ldots n;\eqno (A.9)
\end{equation*}
\begin{equation*}
\frac{\delta \mathcal{A}}{\delta j^i}=\frac{\partial \mathcal{L}}{\partial
j^i}-p_i=0,\qquad i=0,1,\ldots n\eqno (A.10)
\end{equation*}
Variation with respect to $p_i$ leads to Eq.(A.7).

Using identities
\begin{equation*}
\partial _k(\frac{\partial ^2J}{\partial \xi _{0,l}\partial \xi _{i,k}}%
)\equiv 0,\qquad i,l=0,1,\ldots n,\qquad \partial _kj^k=\partial _k\frac{%
\delta J}{\delta \xi _{0,i}}\equiv 0,\eqno (A.11)
\end{equation*}
\begin{equation*}
\frac{\partial ^2J}{\partial \xi _{0,i}\partial \xi _{j,k}}\xi _{j,l}\equiv
\delta _l^k\frac{\partial J}{\partial \xi _{0,i}}-\delta _l^i\frac{\partial J%
}{\partial \xi _{0,k}},\qquad i,k,l=0,1,\ldots n,\eqno (A.12)
\end{equation*}
and eliminating the variables $\xi $, one can reduce equations
(A.7), (A.9), (A.10) to the conventional hydrodynamic form (2.9),
(2.14)

Instead of elimination of variables $\xi $ Eqs. (A.9), (A.10) can be
integrated in the form
\begin{equation*}
p_{k}=\frac{\partial \mathcal{L}(x,j)}{\partial j^{k}}=\hbar \partial
_{k}\varphi +\hbar g^{\beta }(\mathbf{\xi })\partial _{k}\xi _{\beta
};\qquad k=0,1,\ldots n,\eqno(A.13)
\end{equation*}%
where $g^{\beta }(\mathbf{\xi })$, $\beta =1,2,\ldots n$ are arbitrary
functions of the argument $\mathbf{\xi }=\{\xi _{\alpha }\}$, $\alpha
=1,2,\ldots $n. $\varphi $ is some new variable. $\hbar $ is the Planck
constant which is introduced, for variables $\varphi $, $\mathbf{\xi }=\{\xi
_{\alpha }\}$, $\alpha =1,2,\ldots n$ and the functions $g^{\beta }$ were
dimensionless. In this context $\hbar $ has no quantum meaning.

Indeed, let us substitute relation (A.13) into Eq.(A.9) and use the first
Eq.(A.11), antisymmetry of $\partial ^{2}J/\partial \xi _{0,l}\partial \xi
_{\alpha ,k}$ and symmetry of $\partial _{k}\partial _{l}\varphi $ with
respect to indices $l,k$. Then one obtains
\begin{equation*}
\partial _{k}[\frac{\partial ^{2}J}{\partial \xi _{0,l}\partial \xi _{\alpha
,k}}g^{\beta }(\mathbf{\xi })\xi _{\beta ,l}]=0,\qquad \alpha =1,2,\ldots n.%
\eqno(A.14)
\end{equation*}%
Now using the identity
\begin{equation*}
\frac{\partial ^{2}J}{\partial \xi _{0,l}\partial \xi _{\alpha ,k}}\xi
_{\beta ,l}\equiv -\delta _{\beta }^{\alpha }\frac{\partial J}{\partial \xi
_{0,k}},\qquad \alpha ,\beta =1,2\ldots n,\qquad k=0,1,\ldots n,\eqno(A.15)
\end{equation*}%
and the second relation (A.11), one can reduce Eq.(A.14) to the form

\begin{equation*}
-\delta _{\beta }^{\alpha }\frac{\partial J}{\partial \xi _{0,k}}\xi
_{\gamma ,k}{\frac{\partial g^{\beta }(\mathbf{\xi })}{\partial \xi _{\gamma
}}}=0,\qquad \alpha =1,2,\ldots n.\eqno(A.16)
\end{equation*}%
Due to the identity
\begin{equation*}
\frac{\partial J}{\partial \xi _{0,k}}\xi _{l,k}\equiv \delta
_{l}^{0}J,\qquad l=0,1,\ldots n\eqno(A.17)
\end{equation*}%
the equation (A.16) is satisfied for any functions $g^{\beta }(\mathbf{\xi }%
) $, $\beta =1,2,\ldots $n. Thus, the relation (A.13) is a general solution
of the system of equations (A.9).

Let us substitute Eq.(A.13) into the action (A.8) and use that due to
Eq.(A.17)

\begin{equation*}
\frac{\partial J}{\partial \xi _{0,k}}p_k=\hbar \frac{\partial (\varphi ,\xi
_1,\xi _2,\ldots \xi _n)}{\partial (x^0,x^1,\ldots x^n)}.\eqno (A.18)
\end{equation*}
Then the action (A.8) turns to Eq.(2.2). The term (A.18) is omitted, because
it does not give any contribution into dynamic equations. Thus, Eq.(2.2) is
a corollary of Eq.(2.1).

Now let the variational problem with the functional (2.2) take place.
Corresponding dynamic equations have the form (2.9)--(2.11). The relations
(A.7) satisfy Eqs. (2.9), (2.11) identically, and an usage of them as
additional constraints does not change the variational problem. Substituting
(A.7) into Eq.(2.2) and using identities (A.17), (A.18), one obtains
Eq.(A.4) which is equivalent to Eq.(2.1). Thus, the variational problems
with actions (2.1) and (2.2) are equivalent.

\bigskip \centerline{APPENDIX $B$} \bigskip

Let us show here that under conditions
\begin{equation*}
\mathbf{\xi }^{2}=1,\qquad \mathbf{z}^{2}=1\eqno(B.1)
\end{equation*}%
the equation
\begin{equation*}
-\dot{\mathbf{\xi }}\times \mathbf{z}+{\frac{(\mathbf{z}\dot{\mathbf{\xi }})%
}{2(1+\mathbf{z\xi })}}\mathbf{\xi }\times \mathbf{z}+{\frac{\mathbf{\xi }(%
\dot{\mathbf{\xi }}\times \mathbf{z})}{2(1+\mathbf{z\xi })}}\mathbf{z}-{%
\frac{(1+\mathbf{z\xi })}{2}}\mathbf{b}=C\mathbf{\xi },\eqno(B.2)
\end{equation*}%
where $\mathbf{b}$ is a given vector, and $C$ is an indefinite function of $%
\tau $, reduces to the form
\begin{equation*}
\dot{\mathbf{\xi }}=\mathbf{b\times \xi }\eqno(B.3)
\end{equation*}%
Let $\mathbf{z\xi }=\cos \alpha \neq \pm 1.$ Let us introduce an orthonormal
basis $\{\mathbf{e}_{\alpha }\}$, $\alpha =1,2,3$
\begin{equation*}
\mathbf{e}_{3}=\mathbf{\xi },\qquad \mathbf{e}_{2}={\frac{(\mathbf{\xi }%
\times \mathbf{z})}{\sin \alpha }},\qquad \mathbf{\xi }=\cos \alpha \eqno%
(B.4)
\end{equation*}%
\begin{equation*}
\mathbf{e}_{1}=\mathbf{e}_{2}\times \mathbf{e}_{3}={\frac{\mathbf{\xi }%
\times (\mathbf{\xi }\times \mathbf{z})}{\sin \alpha }}={\frac{\mathbf{z}-%
\mathbf{\xi }\cos \alpha }{\sin \alpha }}\eqno(B.5)
\end{equation*}%
Then
\begin{equation*}
\mathbf{z}=\mathbf{e}_{1}\sin \alpha +\mathbf{e}_{3}\cos \alpha ,\qquad \dot{%
\mathbf{\xi }}=\dot{\xi}_{\alpha }\mathbf{e}_{\alpha },\qquad \mathbf{b}%
=b_{\alpha }\mathbf{e}_{\alpha }\eqno(B.6)
\end{equation*}%
It follows from (B.1) and (B.6) that
\begin{equation*}
\mathbf{\xi }\dot{\mathbf{\xi }}=0,\qquad \dot{\xi _{3}}=0,\qquad z_{2}=0.\eqno%
(B.7)
\end{equation*}%
Substituting Eqs.(B.6), (B.7) into Eq.(B.2) and equating
coefficients before the basis vectors $\mathbf{e}_{\alpha }$, one
obtains
\begin{equation*}
\mathbf{e}_{1}:\qquad \dot{\xi}_{2}=-b_{1}\eqno(B.8)
\end{equation*}%
\begin{equation*}
\mathbf{e}_{2}:\qquad \dot{\xi}_{1}=b_{2}\eqno(B.9)
\end{equation*}%
\begin{equation*}
\mathbf{e}_{3}:\qquad C=\frac{b_{3}}{2}(1+\cos \alpha )-\frac{2+\cos \alpha
}{2(1+\cos \alpha )}\sin \alpha \dot{\xi}_{2}.\eqno(B.10)
\end{equation*}%
Eq.(B.3) follows from Eqs.(B.7)-(B.9). Eqs.(B.8)--(B.10) determine the
indefinite function $C$ through the given vector $\mathbf{b}$ and the scalar
product $(\mathbf{z\xi })$.

If $\mathbf{z}=\mathbf{\xi }$, the equation (B.2) reduces to the form
\begin{equation*}
-\dot{\mathbf{\xi }}\times \mathbf{\xi }-\mathbf{b}=C\mathbf{\xi }.\eqno%
(B.11)
\end{equation*}%
Forming the vector product of Eq.(B.11) with $\mathbf{\xi }$ and using
(B.7), one obtains Eq.(B.3).

If $\mathbf{z}=-\mathbf{\xi }$, the equation (B.2) becomes indefinite, and
it is necessary to use the limit $\mathbf{\xi }\rightarrow -\mathbf{z}$. It
leads to the result (B.3).

\medskip \centerline{REFERENCES }

\noindent [1] Yu.A.Rylov, \textit{Physics Essays}. \textbf{4}, 300, (1991).

\noindent [2] Yu.A.Rylov, \textit{J. Math. Phys.}. \textbf{32}, 2092, (1991).

\noindent \lbrack 3] N.N.Bogolyubov, D.V.Shirkov, \textit{Introduction to
Theory of Quantized Fields}, Moscow, Nauka, 1984, (in Russian) chp.1, sec.7.

\noindent [4] F.Sauter, \textit{Zs.Phys.} \textbf{63}, 803, (1930), \textbf{%
64}, 295, (1930).

\noindent [5] A.Sommerfeld, \textit{Atombau and Spektrallinien.} bd.2,
Braunschweig, 1951.

\noindent [6] J.L.Anderson, \textit{Principles of Relativity Physics.}
Academic Press, New York, 1967. pp. 75--88.

\noindent [7] D.Hestenes, \textit{Space-Time Algebra.} Gordon and Breach,
New York, 1966

\noindent [8] D.Hestenes, \textit{Advances in Applied Clifford Algebras}
\textbf{2}, 215, (1992).

\noindent [9] A.O.Barut, N.Zanghi, \textit{Phys. Rev. Lett}. \textbf{52},
2009, (1984).

\noindent [10] E.Schr\"odinger, \textit{Sitzungsber. Preuss. Wiss. Phys.
Math. Kl.} \textbf{24}, 418, (1930).

\noindent [11] A.O.Barut, A.J.Bracken, \textit{Phys. Rev}. \textbf{D23},
2454, (1981).

\noindent [12] J.C.Aron, \textit{Found. Phys.} \textbf{11}, 863, (1981).

\noindent [13] D.Hestenes, \textit{Found. Phys.} \textbf{20}, 45, (1990).

\noindent [14] W.A.Rodriges Jr., J.Vaz Jr. \textit{Phys. Lett.} \textbf{B318}%
, 623, (1993).

\noindent [15] Yu.A.Rylov, \textit{J. Math. Phys.} \textbf{33}, 4220, (1992).


\begin{thebibliography}{9}
\bibitem{R2004} Is the Dirac particle composite? \textit{e-print,
/physics/0410045}.

\bibitem{R2004a} Is the Dirac particle completely relativistic? \textit{%
e-print, /physics/0412032}.
\end{thebibliography}
\end{document}